\documentclass[submission,copyright,creativecommons]{eptcs}
\providecommand{\event}{SYNT 2016} 

\usepackage[final,formats]{listings}
\usepackage{multicol}
\usepackage{graphicx}
\usepackage{wrapfig}
\usepackage{amsfonts}
\usepackage{enumitem}
\usepackage{paralist}
\usepackage{cite}

\let\subparagraph\paragraph
\usepackage{titlesec}
\titlespacing\subsection{0pt}{10pt plus 4pt minus 2pt}{3pt plus 2pt minus 2pt}

\newcommand{\src}[1]{\texttt{#1}}

\newcommand{\mypara}[1]{\vspace{1mm}\noindent\emph{#1}~~}

\newtheorem{lesson}{Lesson}{\it}{\it}
\newtheorem{ex}{Example}{\it}{\it}
\newtheorem{definition}{Definition}{\it}{\it}

\lstnewenvironment{tsllisting}[1][]
{\lstset{
    escapeinside={(*@}{@*)},
    basicstyle=\ttfamily\scriptsize,
    keywordstyle=\bfseries,
    keywordstyle=\bfseries,
    sensitive=false,
    morekeywords={pause, if, else, typedef, task, export, template, endtemplate, process, controllable, forever, wait, return, assert, goal, instance},
    identifierstyle=, 
    commentstyle=\slshape, 
    stringstyle=, showstringspaces=false,
    sensitive=false,
    morecomment=[s]{/*}{*/},
    morecomment=[l]{//},
    numberstyle=\tiny,
    stepnumber=1,
    numbersep=1pt,
    emphstyle=\bfseries,
    belowskip=0pt,
    aboveskip=0pt,
    #1
}}{}

\def\titlerunning{Developing a Practical Reactive Synthesis Tool}
\def\authorrunning{L. Ryzhyk, A. Walker}

\begin{document}

\title{Developing a Practical Reactive Synthesis Tool:\\ 
Experience and Lessons Learned}

\author{Leonid Ryzhyk\footnote{Work completed at the University of 
Toronto}
\institute{Samsung Research America}
\email{l.ryzhyk@samsung.com}
\and
Adam Walker
\institute{NICTA and UNSW, Sydney, Australia}
\email{adamwalker10@gmail.com}
}

\maketitle

\begin{abstract}

    We summarise our experience developing and using Termite, the 
    first reactive synthesis tool intended for use by software 
    development practitioners.  We identify the main barriers to 
    making reactive synthesis accessible to software developers 
    and describe the key features of Termite designed to overcome 
    these barriers, including an imperative C-like specification 
    language, an interactive source-level debugger, and a 
    user-guided code generator.  Based on our experience applying 
    Termite to synthesising real-world reactive software, we 
    identify several caveats of the practical use of the reactive 
    synthesis technology.  We hope that these findings will help 
    define the agenda for future research on practical reactive 
    synthesis.
    

\end{abstract}

\section{Introduction}\label{s:intro}

The reactive synthesis method aims to automatically generate an 
implementation of a reactive system based on a declarative 
specification of its desired behaviour~\cite{Pnueli_Rosner_89}.  
While theoretical and algorithmic aspects of reactive synthesis 
are fairly well understood, there exists virtually no practical 
experience using this technology to synthesise real-world software 
and hardware.  Existing synthesis tools have been developed as 
research vehicles for experimentation with new synthesis 
algorithms rather than as practical tools for use by hardware or 
software 
designers~\cite{Bloem_CGHKRSS_10,Jobstmann_Bloem_06,Jobstmann_GWB_07,Ehlers_10,Bohy_BFJR_12,Cheng_KLB_11}.
The gap between theory and practice hinders further advances in 
the field.  For comparison, in the closely related field of 
automatic verification, the push towards real-world application of 
the model checking technology has fueled revolutionary advances of 
the technology itself~\cite{Grumberg_Veith_08}. 

For the past several years, our team has been working on using 
reactive synthesis to automate the development of operating system  
device drivers.  The primary objective of the project has been to 
improve driver developers' productivity, as opposed to the 
creation of new synthesis algorithms or tools.  However, early on 
in the project we discovered that none of the existing tools were 
suitable for our purposes.  As a result, we created Termite, the 
first synthesis tool designed to make reactive synthesis 
accessible to software developers as a well-defined, predictable 
methodology.

Our previous publications describe the synthesis algorithm of 
Termite~\cite{Walker_Ryzhyk_14} and its application to the driver 
synthesis problem~\cite{Ryzhyk_WKLRSV_14}.  In this paper we focus 
on our experience designing and using a practical user-friendly 
synthesis tool.  Our contributions are three-fold.  First, we 
present the rationale behind the design of a practical reactive 
software synthesis toolkit.  We consider each step involved in the 
synthesis process: specification development, strategy synthesis, 
specification debugging, and code generation, and identify the 
main barriers to making these tasks accessible to software 
developers.

Second, we present the key features of Termite designed to 
overcome these barriers: (1) an imperative C-like specification 
language, which enables developers to apply conventional 
programming techniques in writing specifications for reactive 
synthesis, (2) an interactive source-level debugger, which helps 
the developers to troubleshoot synthesis failures, and (3) a 
user-guided code generator, which combines the power of automation 
with the flexibility of manual development.

Third, and most importantly, we identify several caveats of the 
practical use of the reactive synthesis technology, based on our 
experience applying Termite to synthesising real-world reactive 
software.  We hope that these findings will help define the agenda 
for future research on practical reactive synthesis.  We formulate 
these caveats as ``lessons learned'' throughout the paper.

\section{User-guided synthesis: the user perspective}

Termite is intended for synthesis of reactive software modules 
such as device drivers or robotic controllers.  A reactive module 
is typically embedded in a larger software stack written in C or 
other imperative language.  The module is activated by invoking 
its event handler functions.  It issues control actions via 
callbacks to the environment.  


\begin{ex}[Running example: a jukebox controller]
Consider the control module of a mechanical jukebox.  The jukebox 
consists of (1) a rotating drum holding up to $256$ vinyl records, (2) a 
turntable, (3) a mechanical arm that transfers records from the 
drum to the turntable, and (4) a number pad for choosing the next 
record to play.  The control module controls the jukebox mechanism 
by issuing commands to (1) rotate the drum to a specified 
position, (2) place the record at the current position on the 
turntable, (3) move the record from the turntable back to the 
drum, and (4) play the record on the turntable.  

Figure~\ref{f:synthesised} shows our synthesised implementation of 
the control module with detailed comments.  It consists of four 
event handlers: \src{evt\_selection}, triggered when the user 
selects a record to play, \src{evt\_rotated}, triggered when the 
drum finishes rotation to the requested position, 
\src{evt\_parked}, issued when the mechanical arm finishes 
transferring the record from the drum to the turntable or back, 
and \src{evt\_playback\_complete}, which indicates that the record 
has been played to the end.
\end{ex}

\begin{figure}[t]
\lstset{numbers=left}
\lstset{firstnumber=last}
    \scriptsize
\begin{tsllisting}[multicols=2]
// user has selected a record to play
task void evt_selection() {
  if ((jb.position == jb.selection) && 
      jb.arm_down)
    // selected record already on the
    // turntable-start playing
    jb.cmd_play();
  else if (jb.arm_down)
    // another record is on the 
    // turntable, pick it up first 
    jb.cmd_lift();
  else
    // rotate drum to selected position
    jb.cmd_rotate(jb.selection);
};
// drum has finished rotating
task void evt_rotated() { 
  // place record on the turntable
  jb.cmd_put(); 
};
// mechanical arm has finished moving the 
// record to the turntable or to the drum
task void evt_parked() {
  if(jb.have_selection){
    if (jb.arm_down)
      // record on the turntable-play it
      jb.cmd_play();
    else
      // previous record returned to drum,
      // rotate to the new selection
      jb.cmd_rotate(jb.selection);
  }; 
};
// record has been played to the end
task void evt_playback_complete() { 
  // return it to the drum
  jb.cmd_lift(); 
};
\end{tsllisting}
\vspace{-3mm}
\caption{Synthesised implementation of the jukebox 
controller.}\label{f:synthesised}
\end{figure}

\subsection{Synthesis workflow}\label{s:workflow}

\begin{figure}[t]
    \center
    \includegraphics[width=0.9\linewidth]{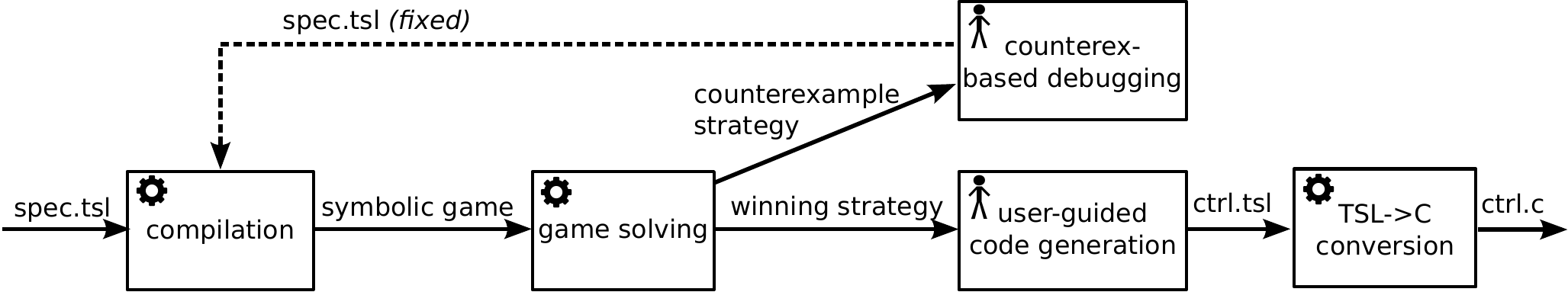}
    \caption{Termite synthesis workflow. The gear and the stick 
    figure distinguish fully automatic and user-guided stages of 
    the workflow.}\label{f:flow}
\end{figure}

The Termite synthesis workflow is shown in Figure~\ref{f:flow}.  
Termite takes as input a specification
in the \emph{Termite Specification Language} (TSL).  The 
specification is compiled into a symbolic 
GR-1~\cite{Piterman_PS_06} (Section~\ref{s:internal}), which is 
then solved by the Termite game solver.  The game solver produces 
either a winning strategy for the controller or, if one does not 
exist, a counterexample strategy on behalf of the environment.  In 
the former case, the winning strategy is converted into a 
controller implementation in TSL in a user-guided fashion.  The 
resulting implementation is automatically translated into a C 
module.  In the latter case, the user explores the counterexample 
strategy using the \emph{Termite visual debugger} in order to 
track the cause of the synthesis failure down to a defect in the 
input specification.  

%

\subsection{The Termite Specification Language}\label{s:tsl}

\mypara{Rationale} Our main language design objective is to 
facilitate the specification and synthesis of a particular class 
of target systems, namely reactive software modules, while 
minimising modelling errors.  We therefore prioritise modelling 
convenience over generality.  In addition to this main principle, 
we identify the following requirements for the language:

\begin{compactenum}
    \item \emph{Modularity.} The ability to model each component 
        of a complex system as a separate module enables 
        specification reuse, which is crucial in practical 
        applications of reactive synthesis, as explained in 
        Section~\ref{s:specs}.
    \item \emph{Programmer-friendly syntax.} TSL is intended for 
        use by software developers rather than formal methods 
        experts.  As such, it must support modelling using
        familiar programming concepts and idioms.
    \item \emph{Modelling of hybrid systems.} In many applications 
        of reactive synthesis, including device driver development 
        and robotics, we deal with systems that contain both 
        software and hardware components.  We therefore designed 
        TSL to facilitate modelling of such hybrid systems.
\end{compactenum}

\mypara{Key features of TSL} Following the first 
requirement, TSL supports the development of modular 
specifications through object-oriented modelling.  
\begin{wrapfigure}[6]{r}{0.23\textwidth}
  \includegraphics[width=0.21\textwidth]{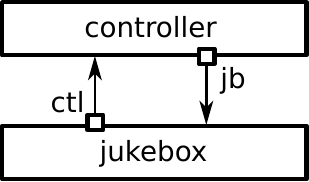}
\end{wrapfigure}
A TSL class, called \emph{template}, consists of \emph{variables}, 
\emph{methods} and \emph{processes}, which together model a 
component of the environment
or the reactive module that is being synthesised.
A TSL specification consists of multiple statically created 
template \emph{instances} that together model a complete system.  
Templates communicate by invoking each other's methods via typed 
interfaces exported through \emph{ports}.
The figure on the right shows the two templates involved in our 
running example.  Arrows represent bindings of template ports.  
Lines~2-3 in Figure~\ref{f:specs} instantiate this architecture in 
TSL.

Following the second requirement, we designed TSL as an imperative 
C-style language (Section~\ref{s:specs}).  
To meet the third requirement, we extend TSL with hardware 
modelling features, such as arbitrary-size bit vectors, bit 
slices, combinatorial logic, etc.  In addition to standard 
programming language features, TSL supports constructs specific to 
reactive synthesis, introduced below.  A complete description of 
TSL syntax can be found in the language reference 
manual~\cite{tsl}.  

\subsection{Developing input specifications}\label{s:specs}

A Termite specification consists of at least two templates, which 
model the environment and the controller.  The environment model 
can be seen as a \emph{test harness} for the controller, which 
stimulates controller inputs and validates its outputs.  TSL 
allows structuring such a model similar to how one would write a 
software test harness: it consists of a workload generator process 
that generates a stream of requests to the controller and callback 
methods invoked by the controller.


The controller template declares methods of the controller and 
specifies the control objective.  Termite supports GR-1 objectives 
(see Piterman et al.~\cite{Piterman_PS_06} and 
Section~\ref{s:internal}), defined in terms of one or more 
\emph{goal conditions}, which the controller must satisfy 
infinitely often.  For example, a goal condition may require the 
controller to eventually complete every request from the 
environment.
In addition, the controller template can provide partial 
implementation of controller methods.  During synthesis, Termite 
tries to extend this to a complete implementation that satisfies 
the objective.  




\begin{ex}
In our running example, the environment consists of the jukebox 
mechanism modelled by the \src{jukebox} template in line~6, 
Figure~\ref{f:specs}.  The mechanism can be in one of the states 
listed in lines~9-13.  Variables in lines~15-23 model the state of 
the number pad, the drum and the mechanical arm.  The main part of 
the model is the process in line~24, which generates a stream of 
events to the controller.
The first part of the loop body (lines~27-32) simulates the user 
selecting the next record to play.  In our simple jukebox, record 
selection is only enabled when there is no current selection yet 
and the jukebox mechanism is idle (line~27).  Even when the event 
is enabled, it is not guaranteed to occur---it is up to the user 
when to make a selection.  We model such external choice in 
line~29, where the asterisk (`$*$') represents a 
non-deterministically chosen value.  Likewise, the record number 
selected by the user is modelled as a non-deterministic value 
passed as an argument to the event handler in line~31.

\begin{figure}[t]
\lstset{numbers=left}
    \scriptsize
\begin{tsllisting}[multicols=2]
template main
  instance controller ctl(jb);
  instance jukebox    jb(ctl);
endtemplate

template jukebox(controller ctl)
  // states of the jukebox mechanism
  typedef enum {
    idle,
    spin,//spinning the drum
    play,//playing a record
    put, //moving record to turntable
    lift //returning record to drum
  } state_t;
  state_t state = idle;//current state
  // record has been selected
  bool have_selection = false;
  // selected record number
  uint8 selection;
  // mechanical arm position
  bool arm_down;
  // current drum position
  uint8 position;
  process pjukebox {
    forever {
      // Simulate user selection
      if (!have_selection &&
          state == idle) {
        if (*) {
          have_selection = true;
          selection = *;
          ctl.evt_selection();};};
      // Simulate command completion
      if (state == spin) {
        state = idle;
        ctl.evt_rotated();
      } else if (state == put) {
        state = idle;
        arm_down = true;
        ctl.evt_parked();
      } else if (state == lift) {
        state = idle;
        arm_down = false;
        ctl.evt_parked();
      } else if (state == play) {
        state = idle;
        have_selection = false;
        ctl.evt_playback_complete();
      };
      pause;
    };
  };
  task controllable void 
  cmd_rotate(uint8 pos) {
    assert(state==idle && !arm_down);
    state = spin;
    position = pos;
  };
  task controllable void cmd_put() {
    assert(state==idle && !arm_down);
    state = put;
  };
  task controllable void cmd_lift() {
    assert(state==idle && arm_down);
    state = lift;
  };
  task controllable void cmd_play() {
    assert(state==idle && arm_down);
    assert(have_selection &&
           (position==selection));
    state = play;
  };
endtemplate

template controller(jukebox jb)
  goal play_selection =
       (jb.have_selection == false);
  task void evt_selection(){...;};
  task void evt_rotated(){...;};
  task void evt_parked(){...;};
  task void evt_playback_complete(){...;};
endtemplate
\end{tsllisting}
\caption{Input TSL specification for the jukebox controller.}\label{f:specs}
\end{figure}

The second part of the loop (lines~34-49) generates a completion 
event for the current operation peformed by the jukebox.  For 
example, if the jukebox is in the \src{put} state, transferring 
the record from the drum to the turntable, it returns to the idle 
state (line~38), updates arm position, and issues the 
\src{evt\_parked} event, which enables the controller to respond 
to the state transition.

Lines~53-72 show environment callbacks that can be invoked by the 
controller, designated by the \src{controllable} qualifier.  Each 
callback contains an assertion that specifies when it is safe to 
issue the command.
For example, the rotate command (line~53) can only be issued when 
the mechanism is idle and the mechanical arm is in the upward 
position (i.e., there is no record on the turntable).

Lines~75-82 show the controller template.  The goal condition in 
line~76 requires that the system is in a state where there is no 
user selection on the number pad.  This can only be achieved by 
playing each selected record.  The bodies of all template methods 
contain elipsis (``\src{...}''), which are placeholders for 
synthesised code, referred to as \emph{magic blocks}.
\end{ex}

Magic blocks are regular TSL statements and can be used anywhere 
in the code.  This enables the developer to enforce a preferred 
implementation structure in advance by provinding a partial 
implementation of a method, for example:
\vspace{1mm}
\begin{tsllisting}[frame=single]
task void evt_selection() {
  if (((jb.position == jb.selection) && jb.arm_down)) {...;}
  else if (jb.arm_down) {...;}
  else {...;};
};
\end{tsllisting}

Multiple TSL statements are combined in a single atomic 
transition.  The transition terminates upon reaching a special 
\src{pause} statement or a magic block.  For example, from its 
initial state in line~25, the \src{pjukebox} process may execute 
atomically until calling the \src{evt\_selection} method in 
line~32, where it stops at the magic block in line~78.  
Alternatively, if none of the conditions in the body of the 
process holds, the transition stops at the pause location in 
line~50.

Note that in this example the specification is longer and more 
complicated than the resulting synthesised controller 
(Figure~\ref{f:synthesised}).  We observed a similar effect in  
synthesising real-world device drivers.
This is not surprising, as artificially engineered systems are 
usually designed to have a simple control strategy.  This is in 
contrast to, for example, competitive games such as checkers, 
where the winning strategy is much more complicated than the 
description of game rules.

In order to obtain a productivity improvement with reactive 
synthesis, one must minimise the effort invested in the 
development of input specifications.  This can be achieved by 
reusing existing specifications where possible.  For example, in 
device driver synthesis, the driver environment consists of the 
operating system (OS) and the hardware  
device~\cite{Ryzhyk_WKLRSV_14}.  The OS specification is developed 
once for each class of drivers (e.g., network drivers) and used to 
synthesise many drivers of this type.  The device specification is 
produced by hardware developers as part of the circuit design 
process and can be reused, with some modifications, in driver 
synthesis.  The  object-oriented nature of TSL comes in handy 
here, as it allows decomposing specifications into reusable 
modules.  

\begin{lesson}
 Specification reuse is essential to achieving a productivity gain 
 with reactive synthesis.
\end{lesson}

\subsection{Counterexample-based debugging}\label{s:debugging}

\mypara{Rationale} The input specification may contain defects 
making it \emph{unrealisable}, i.e., such that there does not 
exist a specification-compliant controller implementation.  In 
this case, Termite produces a certificate of unrealisability in 
the form of a counterexample \emph{spoiling strategy} on behalf of 
the environment.  By following this strategy, the environment is 
able to either force the system into an error state by violating 
one of its assertions or to permanently keep the system from 
entering one of its goal regions.  By studying the counterexample 
strategy, the user can trace the synthesis failure to a bug in the 
specification.  However, such counterexample-based debugging is a 
non-trivial task, which requires special tool support.  The 
counterexample strategy may represent a complex branching 
behaviour that does not have a compact user-readable 
representation.  

\mypara{The interactive debugger} Termite helps the user to 
understand counterexample strategies by exploring them 
interactively.  To this end, the Termite debugger simulates a game 
where the user plays on behalf of the controller, while the 
debugger responds on behalf of the environment, according to the 
counterexample strategy.  The debugger maintains the look and feel 
of a conventional software debugger, where the user steps through 
the code of the system, observing its state and control flow.  
Unlike a conventional debugger, Termite automatically resolves 
environment choices in accordance with the counterexample 
strategy.
At every step of the debugging session, it chooses a TSL process 
to run and assigns concrete values to all non-determimistic 
expressions (`$*$') it evaluates.

When the debugger schedules a process that is currently inside a 
magic block, it allows the user to choose a controller action to 
execute.  In TSL, a controller action corresponds to an invocation 
of one of controllable methods.  The user specifies the action 
that, they believe, represents a correct controller behaviour by 
either typing a line of code or via an interactive dialog.

At some point in the game the user observes an unexpected 
behaviour, e.g., one of user-provided actions does not change the 
state of the system as expected or one of environment actions 
triggers an assertion violation.  Based on this information, the 
user can revise the faulty specification.

\begin{wrapfigure}[11]{r}{0.55\textwidth}
    \center
    \vspace{-5mm}
    \includegraphics[width=0.85\linewidth]{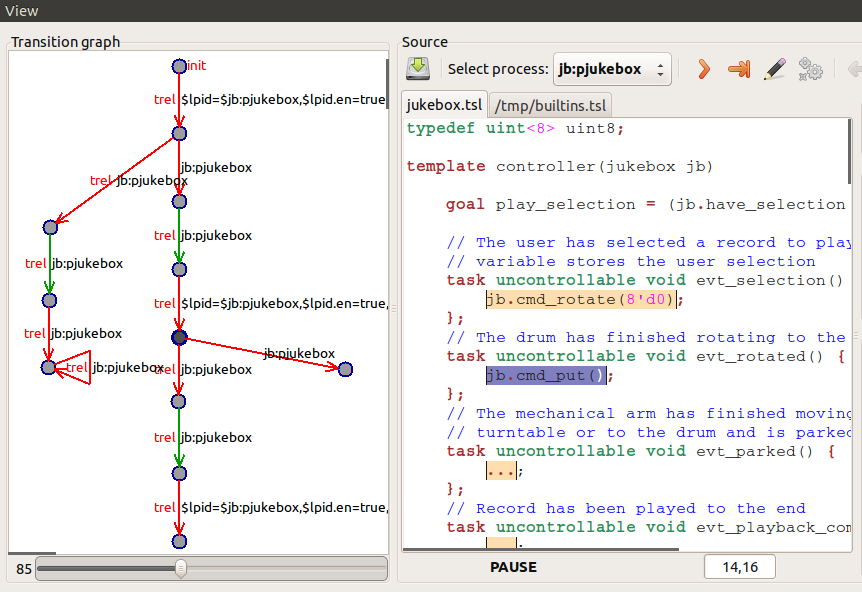}
    \label{f:screenshot}
\end{wrapfigure}
The debugger visualises the transition graph explored during the 
debugging session, as shown in the screenshot on the right.  It 
allows the user to go back to a previously explored state and try 
a different controller behaviour from there.

\begin{ex}
Consider a defective version of our running example where the 
assignment in line~61 is missing.  This is equivalent to the 
jukebox ignoring the command to put the record on the turntable.  
Termite detects that the specification is unrealisable and 
produces a counterexample strategy.  During the debugging session, 
the debugger initialises the \src{arm\_down} variable to false and 
\src{position} to $0$.  It then steps through lines~27-32 of the 
jukebox model, setting the non-deterministic condition in line~29 
to true and the value in line~31 to $0$, thus simulating a request 
to play record number $0$ (i.e., the record in the current 
position).  The debugger stops in the magic block in line~78, 
passing control to the user.   The user attempts to complete the 
request by typing the command \src{jb.cmd\_put()} in the magic 
block to place the record on the turntable.  The user expects the 
environment to respond by invoking the \src{evt\_parked} callback.  
However, due to the specification bug, the \src{(state==put)} 
condition in line~37 does not hold, and the debugger keeps 
iterating the \src{forever} loop without ever invoking any of the 
controller callbacks.  Upon observing this behaviour, the user 
goes back to the invocation of the \src{cmd\_put()} method and 
single-steps its execution, finally discovering that it does not 
set the \src{state} variable as expected.  We have traced the 
synthesis failure to its root cause and can now easily fix it by 
adding the missing statement.
\end{ex}

Our counterexample-guided methodology greatly simplifies 
specification debugging; however it also uncovers an important 
caveat of reactive synthesis: in order to find specification 
defects, the developer must have a good understanding of both the 
input specifications and the expected controller implementation.  

\begin{lesson}
Synthesis does not replace human expertise, as synthesising a 
reactive module requires similar knowledge and skills as 
developing the module manually.
\end{lesson}

Furthermore, the above example shows that debugging synthesis 
failures requires manually providing parts of the controller 
implementation.  In the worst case, the user effectively ends up 
implementing the entire controller manually before finding a bug 
in the specification.  One way to avoid this is to perform 
synthesis \emph{incrementally}.  To this end, we restrict the 
input specification to only exercise a subset of the controller 
functionality (e.g., only initialisation).  This can be achieved 
by commenting out parts of the environment model or by disabling 
all but one goal condition.  Having synthesised the selected 
fragment of the controller, we gradually enable additional 
features until a complete controller has been synthesised.  If at 
some point, the specification becomes unrealisable due to a 
defect, only a modest manual effort is required to locate the 
defect.  Termite supports the incremental synthesis methodology by 
accepting partially implemented controller templates (as described 
in Section~\ref{s:specs}). 

\begin{lesson}
The incremental approach is necessary to streamline specification 
debugging and reduce manual effort involved in synthesis.
\end{lesson}

\subsection{User-guided code generation}\label{s:codegen}

\mypara{Rationale}  Early versions of Termite synthesised the 
reactive module implementation from the input specification fully 
automatically.  Having experimented with this push-button approach 
for several years, we found that it consistently failed to produce 
satisfactory implementations.  First, it proved hard in practice 
to capture all important aspects of the desired behaviour in the 
input specification.  For example, the original version of the 
jukebox controller synthesised by Termite left the last played 
record on the turntable, i.e., it was missing the \src{cmd\_lift} 
command in line~37 in Figure~\ref{f:synthesised}.  This 
implementation satisfies the specification, but delivers 
suboptimal user experience due to an increased delay before 
playing the next selected record.  While in principle the input
specification can be extended to incorporate such additional 
requirements, such extensions tend to be awkward and error-prone.  
In our experience, it is typically easier to achieve the desired 
behaviour via a manual modification to the generated code than to 
enforce it indirectly via changes to the specification.

Second, Termite currently does not handle extensions of the 
reactive synthesis problem, such as quantitative 
synthesis~\cite{Chatterjee_RR_12}, timed 
synthesis~\cite{Cassez_DFLL_05}, and synthesis with imperfect 
information~\cite{Dimitrova_Finkbeiner_12,Raskin_CDH_07}.  As a 
result, it does not automatically enforce non-functional 
properties related to time, performance, power consumption, state 
observability, etc.  This is a deliberate design choice, as these 
extensions make synthesis more computationally expensive and hence 
less practical.  

Third, we found it difficult to achieve a clean, human-readable 
code structure automatically.  This is an important concern, as 
code inspection remains a key quality assurance method even when 
using automatic synthesis.  



\begin{lesson}
Given current state of the art in reactive software synthesis,  
automatic tools are unlikely to replace human expertise in the 
foreseeable future.  A practical synthesis tool must include user 
input in the synthesis process in a way that combines the power 
automation with the flexibility of conventional development.
\end{lesson}

\mypara{User-guided code generation} Our solution to this problem 
is based on the following principles:
\begin{compactenum}
    \item \emph{The two-phase process.} We separate synthesis into 
        a fully automatic \emph{game solving} phase and an 
        interactive \emph{code generation} phase.  The former 
        computes a winning strategy for the controller, which maps 
        states of the system to the set of winning moves in each 
        state, and can be seen as a compact representation of all 
        possible specification-compliant controller 
        implementations.  The latter extracts one specific 
        controller implementation from the strategy in a 
        user-guided fashion.
        
    \item \emph{The user is in control.}  The user interacts with 
        the tool during the code generation phase by arbitrarily 
        changing or amending the generated code.  The automatic 
        code generator is invoked on demand.  It never changes any 
        of the user-provided code, but rather tries to extend it 
        to a complete specification-compliant implementation.

    \item \emph{Enforcing correctness.}  The resulting combination 
        of manual and generated code is continuously validated 
        against the input specification to preserve the strong 
        correctness guarantees offered by automatic synthesis.
\end{compactenum}


\begin{wrapfigure}[10]{r}{0.4\textwidth}
    \center
    \vspace{-4mm}
    \includegraphics[width=0.9\linewidth]{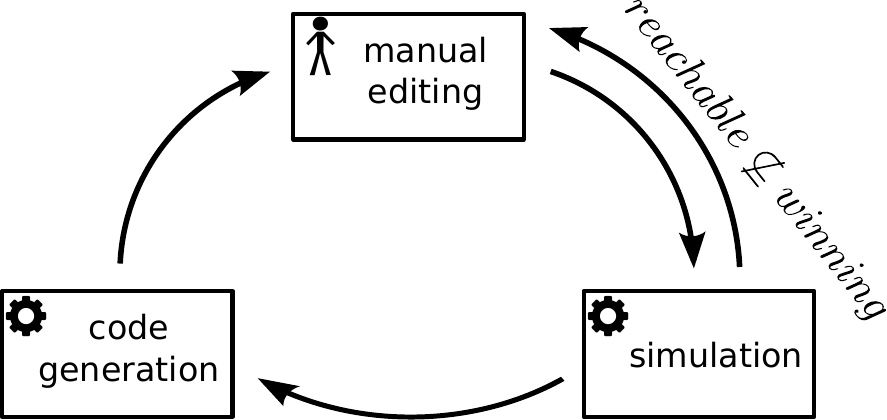}
    \caption{Code generation flow.}\label{f:cgflow}
\end{wrapfigure}
Figure~\ref{f:cgflow} illustrates the interactive code generation 
phase, which iterates over three steps.  At the first step, the 
user edits the partially synthesised implementation inside magic 
blocks (other parts of the module remain read-only).  At the 
moment, Termite does not allow creating new functions 
interactively; however such capability can be readily added in the 
future.  

The manual editing step ends when the user chooses a partially 
filled magic block and invokes the automatic code generator via a 
GUI button to produce the next code statement.  This starts the 
second step, where Termite symbolically simulates execution of the 
system, including all of the manual and automatic code produced so 
far.  The simulation returns a symbolic representation of all 
possible \emph{reachable states} of the system at the control 
location where code generation has been requested.
Termite compares the reachable set against the winning set 
computed during the game solving phase.  If it finds that at least 
one of the reachable states are not winning, this means that the 
partial controller implementation generated so far is incorrect 
due to the manual changes made by the user.
Furthermore, Termite is able to produce a counterexample 
explaining the failure.


During the third step, the Termite code generation algorithm 
translates the winning strategy in the reachable set into a TSL 
statement.  It first checks if there exists a common winning 
action in all reachable states.  Such an action corresponds to a 
simple TSL statement, i.e., an invocation of a controllable 
method. Otherwise, the algorithm computes a partitioning of the 
reachable set into subsets such that there exists a common winning 
action in each subset.  This corresponds to a conditional 
branching statement, such as the one in lines~3-14 in 
Figure~\ref{f:synthesised}.  Finally, the newly generated 
statement is inserted at the corresponding control location, and 
Termite returns to the interactive editing mode.


\begin{ex}
Starting with the empty template, shown in lines~78-81 in 
Figure~\ref{f:specs}, Termite produces code for the 
\src{evt\_selection} and \src{evt\_rotated} methods, which does 
not require any manual changes.  Next, it suggests an empty 
implementation for the \src{evt\_playback\_complete} method.  We 
are not satisfied with this implementation and manually modify it 
to issue the \src{cmd\_lift} command (line~37, 
Figure~\ref{f:synthesised}) in order to return the current record 
to the drum.  Finally, we proceed to generate the 
\src{evt\_parked} method in line~23.  Note that the synthesised 
implementation takes into account manual changes made at the 
previous step.  In particular, the first branch of the generated 
code handles completion of the \src{cmd\_lift} command issued 
manually in line~37.
\end{ex}

\section{Termite internals}\label{s:internal}


In this section we describe the internal design of the Termite 
toolkit, highlighting some of the important implementation issues.

Termite computes a strategy for the controller by representing the 
synthesis problem as a two-player game between the controller and 
the environment.

\begin{definition}[GR-1 game \cite{Piterman_PS_06}]
    \label{d:game}
    Let $F_t(V)$ be a set of Boolean formulas over variables $V$ 
    in some theory $t$ (at the moment, Termite supports the theory 
    of fixed-size bit vectors).  A concrete symbolic GR-1 game 
    $G=\langle X, I, Y_c, Y_u, \delta_c, \delta_u, \Gamma, \Phi 
    \rangle$ consists of a finite set of state variables $X$, an 
    initial condition $I\in F_t(X)$, finite sets $Y_c$ and $Y_u$ 
    of controllable and uncontrollable action variables, 
    controllable transition relation $\delta_c\in F_t(X,Y_c,X')$, 
    uncontrollable transition relation $\delta_u\in F_t(X,Y_u,X')$ 
    (where $X'$ are the next-state versions of state variables 
    $X$),  $\Gamma=(\gamma_1,\ldots,\gamma_k)$, $\gamma_i\in 
    F_t(X)$ is a finite set of goal regions, and $\Phi$ is a 
    finite set of fairness conditions 
    $\Phi=(\phi_1,\ldots,\phi_k)$, $\phi_i\in F_t(X,Y_u)$.
\end{definition}

The controller and the environment participate in the game by 
performing actions allowed by their respective transition 
relations.  To win the game, the controller must infinitely often 
force the system into each goal region $\gamma_i$, assuming fair 
execution where each of the fairness conditions $\phi_j$ holds 
infinitely often.  


The TSL compiler converts a TSL specification into a GR-1 game 
$G$, where state variables $X$ model the state of the TSL program, 
controllable actions encode invocations of controllable methods 
available to the controller, uncontrollable actions model atomic 
transitions of TSL processes
goal sets $\Gamma$ encode goals declared in the input 
specification, and fairness conditions $\Phi$ enforce fair 
scheduling by making sure that every runnable process gets 
scheduled eventually.

We solve the resulting game $G$ using the algorithm by Piterman et 
al.~\cite{Piterman_PS_06}, which involves exhaustive exploration 
of the state space of the game.
The Termite game solver mitigates the state explosion problem by 
constructing and iteratively refining an abstraction of the game 
that approximates states and actions using Boolean predicates.  
Abstraction refinement is performed fully automatically.  
We refer the reader to our earlier 
publication~\cite{Walker_Ryzhyk_14} for a detailed description of 
the abstraction refinement algorithm of Termite.

Throughout the synthesis process, Termite maintains three distinct 
representations of the synthesis problem: (1) a TSL specification, 
(2) a concrete symbolic game, and (3) an abstract game.  All 
interactions with the user, including specification development, 
debugging, and interactive code generation, occur at the TSL 
level.  Internally, Termite completes all user requests at the 
abstract level and lifts results of this computation back to the 
TSL level, which enables it to achieve interactive performance 
during debugging and code generation.  

\begin{figure}[t]
    \center
    \includegraphics[width=0.85\linewidth]{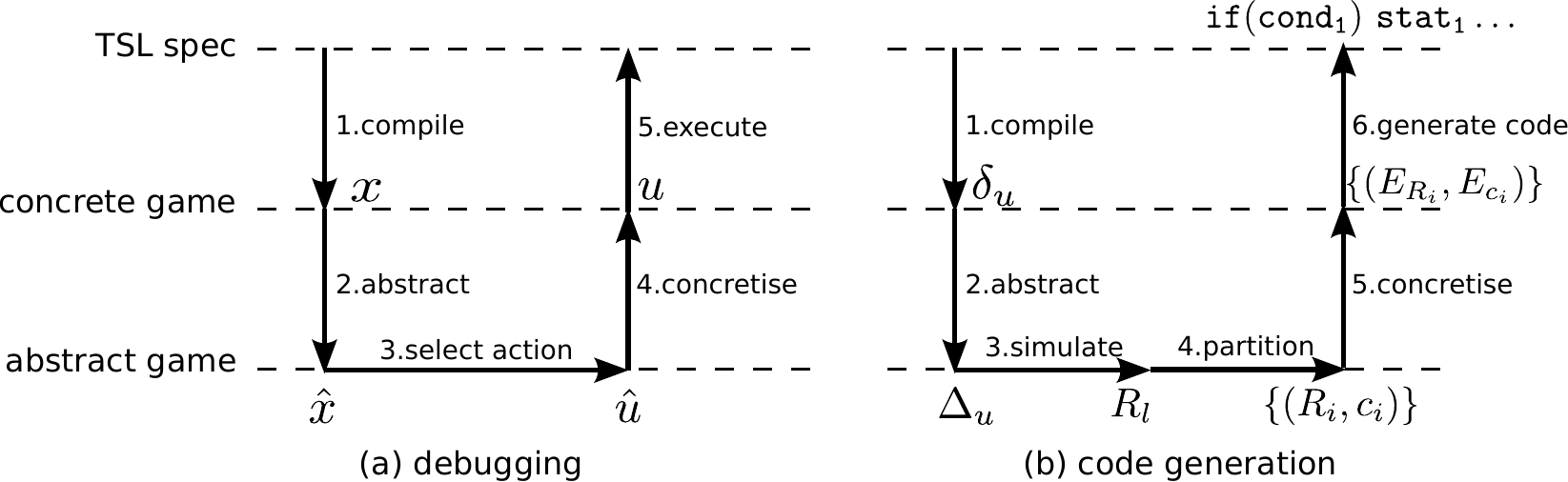}
    \caption{Navigating between different representations of the 
    synthesis problem during debugging and code generation.}
    \label{f:scenario}
\end{figure}

As an example, consider one iteration of a debugging session, 
where the user invokes Termite to pick an uncontrollable action 
according to the counterexample strategy.  It involves the 
following steps (Figure~\ref{f:scenario}a). (1) 
\emph{Compilation:} map the state of the program in the debugger 
to a state $x$ of the underlying game.  (2) \emph{Abstraction:} 
compute corresponding state $\hat{x}$ of the abstract game. (3) 
\emph{Action selection:} pick action $\hat{u}$ from the abstract 
counterexample strategy in state $\hat{x}$. (4) 
\emph{Concretisation:} compute a concrete action $u$ that matches 
the abstract action $\hat{u}$ (there can exist multiple such 
concrete actions). (5) \emph{Execution:} the resulting 
uncontrollable action $u$ encodes the selection of a TSL processes 
to execute, along with choices of all non-deterministic values 
encountered by the process; the user can now simulate this 
transition at the TSL level by single-stepping through its 
statements or by running it to completion.

%
%

As another example, Figure~\ref{f:scenario}b shows one iteration 
of an interactive code generation session, where the user has made 
some changes to the synthesised code and invokes Termite to 
generate a code statement for a selected magic block.  (1) 
\emph{Compilation:} invoke the TSL compiler on the modified 
specification.  All code written or generated so far becomes part 
of the uncontrollable transition relation $\delta_u$ (as Termite 
is not allowed to modify the behaviour of this code). (2) 
\emph{Abstraction:} compute an abstract approximation $\Delta_u$ 
of $\delta_u$.  
(3) \emph{Simulation:} simulate all possible executions of the 
partial implementation generated so far (see 
Section~\ref{s:codegen}): $R=\Delta_u^*(\mathcal{I})$, where 
$\Delta_u^*$ is the transitive closure of $\Delta_u$, 
$\mathcal{I}$ is the abstract initial state, and $R$ is the set of 
reachable abstract states of the system.  Restrict $R$ to a subset 
$R_l\subseteq R$ at the control location $l$ where code generation 
has been requested.
(4) \emph{Partitioning:} compute a partitioning 
$R_l=\dot{\bigcup}_i R_i$ of $R_l$, such that for each partition 
$R_i$, there exists a common winning abstract controllable action 
$c_i$.
(5) \emph{Concretisation:} convert each pair $(R_i, c_i)$ into an 
expression $E_{R_i}(X)$ and $E_{c_i}(X, Y_c)$.  These expressions 
encode a subset of concrete states and the concrete action to be 
performed when in one of these states.
(6) \emph{Code generation:} convert each $E_{R_i}$ into a TSL 
expression $\mathtt{cond_i}$ and each $E_{c_i}$ into a TSL 
statement $\mathtt{stat_i}$. Generate a conditional TSL statement 
of the form: $\mathtt{if (cond_1)~stat_1~else~if (cond_2)~stat_2 
\ldots else~stat_n}$.


\section{Evaluation}\label{s:eval}

We implemented all components of the Termite toolkit, including 
the TSL compiler, the game solver, the counterexample debugger, 
the user-guided code generator, and the TSL-to-C compiler, in 
Haskell.  Our implementation uses the CUDD BDD library for 
efficient symbolic manipulations over Boolean relations, and the 
Z3 SMT solver for satisfiability queries over the theory of bit 
vectors, as described in~\cite{Walker_Ryzhyk_14}.  Termite is 
publicly available under the BSD license and can be downloaded 
from the project website \texttt{\url{http://termite2.org}}.

We evaluate Termite by synthesizing drivers for eight I/O devices: 
a UVC-compliant USB webcam, the 16550 UART serial controller, the 
DS12887 real-time clock, the IDE disk controller for Linux, as 
well as seL4 drivers for I2C, SPI, and UART controllers on the 
Samsung exynos 5 chipset 2 and SPI controller on the STM32F10 
chipset\footnote{Our case studies are discussed in detail 
in~\cite{Ryzhyk_WKLRSV_14}.}.  These devices are representative of 
simple peripherals found in many embedded platforms.  The main 
barrier to synthesizing drivers for more advanced devices, e.g., 
high-performance network controllers, is the current lack of 
support for synthesis of direct memory access (DMA) code in 
Termite.  Such code cannot be generated using only reactive 
synthesis techniques and requires support for synthesis of 
pointer-based data structures such as nested lists, circular 
packet buffers, etc.

Table~\ref{t:results} summarises our case studies.  It shows the 
size (in lines of code) of the input TSL specification, the 
complexity of each benchmark in terms of the number of bit-vector 
variables in the GR-1 game and the total number of bits in these 
variables, the time the game solver took to compute a winning 
strategy for the controller, and the size (in lines of TSL code) 
of the synthesised implementation.  As discussed in 
Section~\ref{s:specs}, input specifications in our case studies 
are larger than the synthesised implementations.  We are able to 
mitigate the problem by decomposing the specification into 
reusable device and OS models.

\begin{table}[t]
    \scriptsize
    \center
    \begin{tabular}{|l|c|c|c|c|}
        \hline
                     & input spec(loc)   & vars(bits)   & synt. time(s) & synthesised code (loc)\\
        \hline
        \hline
        webcam       & 487               & 128 (125565) & 215            & 113 \\
        16450 UART   & 289               & 81 (407)     & 210            & 74  \\
        exynos UART  & 380               & 80 (1185)    & 645            & 37  \\
        STM SPI      & 317               & 68 (389)     & 67             & 24  \\
        exynos SPI   & 327               & 83 (933)     & 25             & 40  \\
        exynos I2C   & 326               & 65 (303)     & 45             & 79  \\
        RT clock     & 370               & 92 (810)     & 56             & 84  \\
        IDE          & 668               & 114 (1333)   & 285            & 94  \\
        \hline
    \end{tabular}
    \caption{Summary of case studies.}
    \label{t:results}
\end{table}

\mypara{Performance} 
As can be seen from Table~\ref{t:results}, Termite is able to 
solve games with hunderds to thousands of state variables.  These 
results illustrate the effectiveness of our predicate abstraction 
algorithm combined with the use of BDDs for symbolic game solving.  
However, the battle for scalable reactive synthesis is far from 
over.  Complex I/O devices can have hundreds of configuration 
registers, which translates to tens of thousands of state bits, 
which is beyond the reach of all existing synthesis algorithms, 
including those used in Termite.  We hope that ongoing research 
into scalable reactive synthesis algorithms will help to close 
this gap.


%

\mypara{Interactive debugging}  We found counterexample-based 
debugging to be crucial to streamlining the synthesis process.  
Before the debugger was available, we relied on code inspection 
for troubleshooting synthesis failures, which proved to be an 
unpredictably long process.  The Termite debugger streamlines this 
process, giving us the confidence that any failure can be 
localised by following well-defined steps. A typical debugging 
session takes a few minutes and involves entering only a few 
commands manually before the defect is localised.


\mypara{User-guided code generation} In our case studies, 60\% to 
90\% of the code was generated fully automatically, with the rest 
produced in a user-guided fashion.  The main sources of manual 
changes were (1) suboptimal performance or structure of the 
auto-generated code, and (2) handling of partially observable 
state.  The latter problem occurs in device drivers, as the driver 
strategy often depends on the internal state of the hardware 
device, which is not directly observable by the driver.  As a 
result, Termite generates invalid code that directly accesses 
device-internal variables.  We modified Termite to report such 
situations to the user, prompting them to provide a functionally 
equivalent valid implementation.  For example, whenever the 
auto-generated code accesses a device-internal register, it may be 
possible to obtain the value of the register by issuing an 
additional command to the device.

\section{Related work}\label{s:related}



While our work is the first to address the quality of synthesised 
code in reactive software synthesis, in the adjacent field of 
hardware synthesis significant progress has been made on the 
circuit minimisation problem, concerned with converting the 
synthesised strategy into a circuit with minimal number of 
gates~\cite{Bloem_EKKL_14,Ehlers_KH_12}.

Previous reactive synthesis tools support specification languages 
designed to be as simple and expressive as possible.  For example, 
Unbeast~\cite{Ehlers_10} and Acacia+~\cite{Bohy_BFJR_12} support 
synthesis from LTL specifications, while 
RATSY~\cite{Bloem_CGHKRSS_10} handles specifications written as 
deterministic B\"uchi automata.  Furthermore, many of these 
languages are geared towards hardware synthesis.  One example is 
the Extended AIGER format used in the reactive synthesis 
competition~\cite{Jacobs_14}.  In contrast, TSL is the first 
domain-specific language for reactive \emph{software} synthesis.


Counterexample-based debugging is a standard method of 
troubleshooting synthesis failures.  Anzu~\cite{Koninghofer_HB_09} 
and RATSY~\cite{Bloem_CGHKRSS_10} implement an interactive version 
of this technique.  Another line of work explores ways to produce 
minimal counterexamples that help locate the bug 
faster~\cite{Konighofer_HB_13}.  Termite complements these
techniques by lifting counterexample-based debugging to the source 
code level.

Techniques presented in this paper are independent of the choice 
of the underlying synthesis algorithm and will benefit from any 
future advances in this area. In particular, recent research has 
proposed efficient symbolic BDD and SAT-based synthesis 
algorithms~\cite{Piterman_PS_06,Bloem2014,Morgenstern_GS_13,Narodytska_LBRW_14}.  
Another line of work explores the use of abstraction to mitigate 
the state explosion 
problem~\cite{Alfaro_GJ_04,Alfaro_Roy_07,Dimitrova_Finkbeiner_12,Henzinger_JM_03}.  

The idea of automatically completing partially specified programs 
has been explored in syntax-guided and functional 
synthesis~\cite{Kneuss_KKS_13,Solar-Lezama_RBE_05}, as well as in 
LTL synthesis~\cite{Naujokat_LS_12}.

\section{Conclusion}

While push-button software synthesis may not be feasible in the 
short-term perspective, we argue that developers can take 
advantage of the reactive synthesis technology by combining the 
power of automation with the flexibility of manual development.  
We presented the design and implementation of the first synthesis 
tool that enables such combination via three new techniques: 
user-guided code generation, code-centric interface, and 
interactive source-level debugging.

\nocite{*}
\bibliographystyle{eptcs}
\bibliography{extra}

\begin{thebibliography}{10}
\providecommand{\bibitemdeclare}[2]{}
\providecommand{\surnamestart}{}
\providecommand{\surnameend}{}
\providecommand{\urlprefix}{Available at }
\providecommand{\url}[1]{\texttt{#1}}
\providecommand{\href}[2]{\texttt{#2}}
\providecommand{\urlalt}[2]{\href{#1}{#2}}
\providecommand{\doi}[1]{doi:\urlalt{http://dx.doi.org/#1}{#1}}
\providecommand{\bibinfo}[2]{#2}

\bibitemdeclare{inproceedings}{Alfaro_GJ_04}
\bibitem{Alfaro_GJ_04}
\bibinfo{author}{Luca \surnamestart de~Alfaro\surnameend},
  \bibinfo{author}{Patrice \surnamestart Godefroid\surnameend} \&
  \bibinfo{author}{Radha \surnamestart Jagadeesan\surnameend}
  (\bibinfo{year}{2004}): \emph{\bibinfo{title}{Three-valued abstractions of
  games: uncertainty, but with precision}}.
\newblock In: {\sl \bibinfo{booktitle}{LICS}}, \bibinfo{address}{Turku,
  Finland}, pp. \bibinfo{pages}{170--179}.
\doi{10.1109/lics.2004.1319611}.

\bibitemdeclare{inproceedings}{Alfaro_Roy_07}
\bibitem{Alfaro_Roy_07}
\bibinfo{author}{Luca \surnamestart de~Alfaro\surnameend} \&
  \bibinfo{author}{Pritam \surnamestart Roy\surnameend} (\bibinfo{year}{2007}):
  \emph{\bibinfo{title}{Solving Games Via Three-Valued Abstraction
  Refinement}}.
\newblock In: {\sl \bibinfo{booktitle}{CONCUR}}, \bibinfo{address}{Lisboa,
  Portugal}, pp. \bibinfo{pages}{74--89}.
\doi{10.1007/978-3-540-74407-8\_6}.

\bibitemdeclare{article}{Alur_BHQR_01}
\bibitem{Alur_BHQR_01}
\bibinfo{author}{R.~\surnamestart Alur\surnameend}, \bibinfo{author}{R.~K.
  \surnamestart Brayton\surnameend}, \bibinfo{author}{T.~A. \surnamestart
  Henzinger\surnameend}, \bibinfo{author}{S.~\surnamestart Qadeer\surnameend}
  \& \bibinfo{author}{S.~K. \surnamestart Rajamani\surnameend}
  (\bibinfo{year}{2001}): \emph{\bibinfo{title}{Partial-Order Reduction in
  Symbolic State-Space Exploration}}.
\newblock {\sl \bibinfo{journal}{Formal Methods in System Design}}
  \bibinfo{volume}{18}(\bibinfo{number}{2}), pp. \bibinfo{pages}{97--116}.
\doi{10.1023/A:1008767206905}.

\bibitemdeclare{article}{Amani_CDLORZ_14}
\bibitem{Amani_CDLORZ_14}
\bibinfo{author}{Sidney \surnamestart Amani\surnameend}, \bibinfo{author}{Peter
  \surnamestart Chubb\surnameend}, \bibinfo{author}{Alastair \surnamestart
  Donaldson\surnameend}, \bibinfo{author}{Alexander \surnamestart
  Legg\surnameend}, \bibinfo{author}{Keng~Chai \surnamestart Ong\surnameend},
  \bibinfo{author}{Leonid \surnamestart Ryzhyk\surnameend} \&
  \bibinfo{author}{Yanjin \surnamestart Zhu\surnameend} (\bibinfo{year}{2014}):
  \emph{\bibinfo{title}{Automatic Verification of Active Device Drivers}}.
\newblock {\sl \bibinfo{journal}{{ACM} Operating Systems Review}}
  \bibinfo{volume}{48}(\bibinfo{number}{1}).
\doi{10.1145/2626401.2626424}.

\bibitemdeclare{inproceedings}{Ball_BCLLMORU_06}
\bibitem{Ball_BCLLMORU_06}
\bibinfo{author}{Thomas \surnamestart Ball\surnameend}, \bibinfo{author}{Ella
  \surnamestart Bounimova\surnameend}, \bibinfo{author}{Byron \surnamestart
  Cook\surnameend}, \bibinfo{author}{Vladimir \surnamestart Levin\surnameend},
  \bibinfo{author}{Jakob \surnamestart Lichtenberg\surnameend},
  \bibinfo{author}{Con \surnamestart McGarvey\surnameend},
  \bibinfo{author}{Bohus \surnamestart Ondrusek\surnameend},
  \bibinfo{author}{Sriram~K. \surnamestart Rajamani\surnameend} \&
  \bibinfo{author}{Abdullah \surnamestart Ustuner\surnameend}
  (\bibinfo{year}{2006}): \emph{\bibinfo{title}{Thorough Static Analysis of
  Device Drivers}}.
\newblock In: {\sl \bibinfo{booktitle}{1st EuroSys Conference}},
  \bibinfo{address}{Leuven, Belgium}, pp. \bibinfo{pages}{73--85}.
\doi{10.1145/1217935.1217943}.

\bibitemdeclare{inproceedings}{Bloem_CGHKRSS_10}
\bibitem{Bloem_CGHKRSS_10}
\bibinfo{author}{Roderick \surnamestart Bloem\surnameend},
  \bibinfo{author}{Alessandro \surnamestart Cimatti\surnameend},
  \bibinfo{author}{Karin \surnamestart Greimel\surnameend},
  \bibinfo{author}{Georg \surnamestart Hofferek\surnameend},
  \bibinfo{author}{Robert \surnamestart K\"{o}nighofer\surnameend},
  \bibinfo{author}{Marco \surnamestart Roveri\surnameend},
  \bibinfo{author}{Viktor \surnamestart Schuppan\surnameend} \&
  \bibinfo{author}{Richard \surnamestart Seeber\surnameend}
  (\bibinfo{year}{2010}): \emph{\bibinfo{title}{{RATSY}--a New Requirements
  Analysis Tool with Synthesis}}.
\newblock In: {\sl \bibinfo{booktitle}{CAV}}, \bibinfo{address}{Edinburgh, UK},
  pp. \bibinfo{pages}{425--429}.
\doi{10.1007/978-3-642-14295-6\_37}.

\bibitemdeclare{inproceedings}{Bloem_EKKL_14}
\bibitem{Bloem_EKKL_14}
\bibinfo{author}{Roderick \surnamestart Bloem\surnameend}, \bibinfo{author}{Uwe
  \surnamestart Egly\surnameend}, \bibinfo{author}{Patrick \surnamestart
  Klampfl\surnameend}, \bibinfo{author}{Robert \surnamestart
  Koenighofer\surnameend} \& \bibinfo{author}{Florian \surnamestart
  Lonsing\surnameend} (\bibinfo{year}{2014}): \emph{\bibinfo{title}{SAT-Based
  Methods for Circuit Synthesis}}.
\newblock In: {\sl \bibinfo{booktitle}{FMCAD}}, \bibinfo{address}{Lausanne,
  Switzerland}.
\doi{10.1109/FMCAD.2014.6987592}.

\bibitemdeclare{article}{Bloem_KS_13}
\bibitem{Bloem_KS_13}
\bibinfo{author}{Roderick \surnamestart Bloem\surnameend},
  \bibinfo{author}{Robert \surnamestart K{\"o}nighofer\surnameend} \&
  \bibinfo{author}{Martina \surnamestart Seidl\surnameend}
  (\bibinfo{year}{2013}): \emph{\bibinfo{title}{{SAT}-Based Synthesis Methods
  for Safety Specs}}.
\newblock {\sl \bibinfo{journal}{CoRR}} \bibinfo{volume}{abs/1311.3530}.

\bibitemdeclare{inproceedings}{Bloem2014}
\bibitem{Bloem2014}
\bibinfo{author}{Roderick \surnamestart Bloem\surnameend},
  \bibinfo{author}{Robert \surnamestart K{\"o}nighofer\surnameend} \&
  \bibinfo{author}{Martina \surnamestart Seidl\surnameend}
  (\bibinfo{year}{2014}): \emph{\bibinfo{title}{{SAT}-Based Synthesis Methods
  for Safety Specs}}.
\newblock In: {\sl \bibinfo{booktitle}{VMCAI'14}}, \bibinfo{publisher}{Springer
  Berlin Heidelberg}, \bibinfo{address}{San Diego, CA, USA}, pp.
  \bibinfo{pages}{1--20}.
\doi{10.1007/978-3-642-54013-4\_1}.

\bibitemdeclare{inproceedings}{Bohy_BFJR_12}
\bibitem{Bohy_BFJR_12}
\bibinfo{author}{Aaron \surnamestart Bohy\surnameend},
  \bibinfo{author}{V{\'{e}}ronique \surnamestart Bruy{\`{e}}re\surnameend},
  \bibinfo{author}{Emmanuel \surnamestart Filiot\surnameend},
  \bibinfo{author}{Naiyong \surnamestart Jin\surnameend} \&
  \bibinfo{author}{Jean{-}Fran{\c{c}}ois \surnamestart Raskin\surnameend}
  (\bibinfo{year}{2012}): \emph{\bibinfo{title}{Acacia+, a Tool for {LTL}
  Synthesis}}.
\newblock In: {\sl \bibinfo{booktitle}{CAV}}, \bibinfo{address}{Berkeley,
  California, USA}, pp. \bibinfo{pages}{652--657}.
\doi{10.1007/978-3-642-31424-7\_45}.

\bibitemdeclare{article}{Bryant_86}
\bibitem{Bryant_86}
\bibinfo{author}{Randal~E. \surnamestart Bryant\surnameend}
  (\bibinfo{year}{1986}): \emph{\bibinfo{title}{Graph-Based Algorithms for
  Boolean Function Manipulation}}.
\newblock {\sl \bibinfo{journal}{IEEE Transactions on Computers}}
  \bibinfo{volume}{35}, pp. \bibinfo{pages}{677--691}.
\doi{10.1109/TC.1986.1676819}.

\bibitemdeclare{inproceedings}{Cai_Gajski_03}
\bibitem{Cai_Gajski_03}
\bibinfo{author}{Lukai \surnamestart Cai\surnameend} \& \bibinfo{author}{Daniel
  \surnamestart Gajski\surnameend} (\bibinfo{year}{2003}):
  \emph{\bibinfo{title}{Transaction level modeling: an overview}}.
\newblock In: {\sl \bibinfo{booktitle}{"1st International Conference on
  Hardware/Software Codesign and System Synthesis"}}, \bibinfo{address}{Newport
  Beach, CA, USA}, pp. \bibinfo{pages}{19--24}.
\doi{10.1145/944650.944651}.

\bibitemdeclare{inproceedings}{Case_BM_06}
\bibitem{Case_BM_06}
\bibinfo{author}{Michael~L. \surnamestart Case\surnameend},
  \bibinfo{author}{Alan \surnamestart Mishchenko\surnameend} \&
  \bibinfo{author}{Robert~K. \surnamestart Brayton\surnameend}
  (\bibinfo{year}{2006}): \emph{\bibinfo{title}{Inductively Finding a Reachable
  State Space Over-Approximation}}.
\newblock In: {\sl \bibinfo{booktitle}{Proceedings of the 15th International
  Workshop on Logic and Synthesis}}.


\bibitemdeclare{inproceedings}{Cassez_DFLL_05}
\bibitem{Cassez_DFLL_05}
\bibinfo{author}{Franck \surnamestart Cassez\surnameend},
  \bibinfo{author}{Alexandre \surnamestart David\surnameend},
  \bibinfo{author}{Emmanuel \surnamestart Fleury\surnameend},
  \bibinfo{author}{Kim~G. \surnamestart Larsen\surnameend} \&
  \bibinfo{author}{Didier \surnamestart Lime\surnameend}
  (\bibinfo{year}{2005}): \emph{\bibinfo{title}{Efficient On-the-fly Algorithms
  for the Analysis of Timed Games}}.
\newblock In: {\sl \bibinfo{booktitle}{CONCUR}}, \bibinfo{address}{San
  Francisco, CA, USA}, pp. \bibinfo{pages}{66--80}.
\doi{10.1007/11539452\_9}.

\bibitemdeclare{inproceedings}{Cerny_HRRT_13}
\bibitem{Cerny_HRRT_13}
\bibinfo{author}{Pavol \surnamestart Cerny\surnameend}, \bibinfo{author}{Thomas
  \surnamestart Henzinger\surnameend}, \bibinfo{author}{Arjun \surnamestart
  Radhakrishna\surnameend}, \bibinfo{author}{Leonid \surnamestart
  Ryzhyk\surnameend} \& \bibinfo{author}{Thorsten \surnamestart
  Tarrach\surnameend} (\bibinfo{year}{2013}): \emph{\bibinfo{title}{Efficient
  synthesis for concurrency by semantics-preserving transformations}}.
\newblock In: {\sl \bibinfo{booktitle}{CAV}}, \bibinfo{address}{Saint
  Petersburg, Russia}.
\doi{10.1007/978-3-642-39799-8\_68}.

\bibitemdeclare{inproceedings}{Cerny_HRRT_14}
\bibitem{Cerny_HRRT_14}
\bibinfo{author}{Pavol \surnamestart Cerny\surnameend}, \bibinfo{author}{Thomas
  \surnamestart Henzinger\surnameend}, \bibinfo{author}{Arjun \surnamestart
  Radhakrishna\surnameend}, \bibinfo{author}{Leonid \surnamestart
  Ryzhyk\surnameend} \& \bibinfo{author}{Thorsten \surnamestart
  Tarrach\surnameend} (\bibinfo{year}{2014}):
  \emph{\bibinfo{title}{Regression-free synthesis for concurrency}}.
\newblock In: {\sl \bibinfo{booktitle}{CAV}}, \bibinfo{address}{Vienna,
  Austria}.
\doi{10.1007/978-3-319-08867-9\_38}.

\bibitemdeclare{inproceedings}{Chatterjee_RR_12}
\bibitem{Chatterjee_RR_12}
\bibinfo{author}{Krishnendu \surnamestart Chatterjee\surnameend},
  \bibinfo{author}{Mickael \surnamestart Randour\surnameend} \&
  \bibinfo{author}{Jean-Fran\c{c}ois \surnamestart Raskin\surnameend}
  (\bibinfo{year}{2012}): \emph{\bibinfo{title}{Strategy Synthesis for
  Multi-dimensional Quantitative Objectives}}.
\newblock In: {\sl \bibinfo{booktitle}{CONCUR}}, \bibinfo{address}{Newcastle,
  UK}, pp. \bibinfo{pages}{115--131}.
\doi{10.1007/978-3-642-32940-1\_10}.

\bibitemdeclare{inproceedings}{Chen_MK_07}
\bibitem{Chen_MK_07}
\bibinfo{author}{Mingsong \surnamestart Chen\surnameend},
  \bibinfo{author}{Prabhat \surnamestart Mishra\surnameend} \&
  \bibinfo{author}{Dhrubajyoti \surnamestart Kalita\surnameend}
  (\bibinfo{year}{2007}): \emph{\bibinfo{title}{Towards {RTL} test generation
  from {SystemC} {TLM} specifications}}.
\newblock In: {\sl \bibinfo{booktitle}{HLDVT'07}}, pp. \bibinfo{pages}{91--96}.
\doi{10.1109/hldvt.2007.4392793}.

\bibitemdeclare{inproceedings}{Cheng_KLB_11}
\bibitem{Cheng_KLB_11}
\bibinfo{author}{Chih{-}Hong \surnamestart Cheng\surnameend},
  \bibinfo{author}{Alois \surnamestart Knoll\surnameend},
  \bibinfo{author}{Michael \surnamestart Luttenberger\surnameend} \&
  \bibinfo{author}{Christian \surnamestart Buckl\surnameend}
  (\bibinfo{year}{2011}): \emph{\bibinfo{title}{{GAVS+:} An Open Platform for
  the Research of Algorithmic Game Solving}}.
\newblock In: {\sl \bibinfo{booktitle}{TACAS}},
  \bibinfo{address}{Saarbr\"ucken, Germany}, pp. \bibinfo{pages}{258--261}.
\doi{10.1007/978-3-642-19835-9\_22}.

\bibitemdeclare{article}{Chipounov_KC_12}
\bibitem{Chipounov_KC_12}
\bibinfo{author}{Vitaly \surnamestart Chipounov\surnameend},
  \bibinfo{author}{Volodymyr \surnamestart Kuznetsov\surnameend} \&
  \bibinfo{author}{George \surnamestart Candea\surnameend}
  (\bibinfo{year}{2012}): \emph{\bibinfo{title}{The {S2E} Platform: Design,
  Implementation, and Applications}}.
\newblock {\sl \bibinfo{journal}{ACM Transactions on Computer Systems}}
  \bibinfo{volume}{30}(\bibinfo{number}{1}), pp. \bibinfo{pages}{2:1--2:49}.
\doi{10.1145/2110356.2110358}.

\bibitemdeclare{inproceedings}{Chou_YCHE_01}
\bibitem{Chou_YCHE_01}
\bibinfo{author}{Andy \surnamestart Chou\surnameend}, \bibinfo{author}{Jun-Feng
  \surnamestart Yang\surnameend}, \bibinfo{author}{Benjamin \surnamestart
  Chelf\surnameend}, \bibinfo{author}{Seth \surnamestart Hallem\surnameend} \&
  \bibinfo{author}{Dawson \surnamestart Engler\surnameend}
  (\bibinfo{year}{2001}): \emph{\bibinfo{title}{An Empirical Study of Operating
  Systems Errors}}.
\newblock In: {\sl \bibinfo{booktitle}{18th {ACM} Symposium on Operating
  Systems Principles}}, \bibinfo{address}{Lake Louise, Alta, Canada}, pp.
  \bibinfo{pages}{73--88}.
\doi{10.1145/502059.502042}.

\bibitemdeclare{article}{Clarke_GJLV_03}
\bibitem{Clarke_GJLV_03}
\bibinfo{author}{Edmund \surnamestart Clarke\surnameend}, \bibinfo{author}{Orna
  \surnamestart Grumberg\surnameend}, \bibinfo{author}{Somesh \surnamestart
  Jha\surnameend}, \bibinfo{author}{Yuan \surnamestart Lu\surnameend} \&
  \bibinfo{author}{Helmut \surnamestart Veith\surnameend}
  (\bibinfo{year}{2003}): \emph{\bibinfo{title}{Counterexample-guided
  abstraction refinement for symbolic model checking}}.
\newblock {\sl \bibinfo{journal}{Journal of the ACM}} \bibinfo{volume}{50}, pp.
  \bibinfo{pages}{752--794}.
\doi{10.1145/876638.876643}.

\bibitemdeclare{article}{Clarke_KSY_04}
\bibitem{Clarke_KSY_04}
\bibinfo{author}{Edmund~M. \surnamestart Clarke\surnameend},
  \bibinfo{author}{Daniel \surnamestart Kroening\surnameend},
  \bibinfo{author}{Natasha \surnamestart Sharygina\surnameend} \&
  \bibinfo{author}{Karen \surnamestart Yorav\surnameend}
  (\bibinfo{year}{2004}): \emph{\bibinfo{title}{Predicate Abstraction of
  {ANSI-C} Programs Using {SAT}}}.
\newblock {\sl \bibinfo{journal}{Formal Methods in System Design}}
  \bibinfo{volume}{25}(\bibinfo{number}{2-3}), pp. \bibinfo{pages}{105--127}.
\doi{10.1023/B:FORM.0000040025.89719.f3}.

\bibitemdeclare{inproceedings}{Desai_GJQRZ_13}
\bibitem{Desai_GJQRZ_13}
\bibinfo{author}{Ankush \surnamestart Desai\surnameend}, \bibinfo{author}{Vivek
  \surnamestart Gupta\surnameend}, \bibinfo{author}{Ethan \surnamestart
  Jackson\surnameend}, \bibinfo{author}{Shaz \surnamestart Qadeer\surnameend},
  \bibinfo{author}{Sriram \surnamestart Rajamani\surnameend} \&
  \bibinfo{author}{Damien \surnamestart Zufferey\surnameend}
  (\bibinfo{year}{2013}): \emph{\bibinfo{title}{{P}: safe asynchronous
  event-driven programming}}.
\newblock In: {\sl \bibinfo{booktitle}{34th annual ACM SIGPLAN conference on
  Programming Language Design and Implementation}}, \bibinfo{address}{Seattle,
  Washington, USA}, pp. \bibinfo{pages}{321--332}.
\doi{10.1145/2491956.2462184}.

\bibitemdeclare{inproceedings}{Dimitrova_Finkbeiner_08}
\bibitem{Dimitrova_Finkbeiner_08}
\bibinfo{author}{Rayna \surnamestart Dimitrova\surnameend} \&
  \bibinfo{author}{Bernd \surnamestart Finkbeiner\surnameend}
  (\bibinfo{year}{2008}): \emph{\bibinfo{title}{Abstraction Refinement for
  Games with Incomplete Information}}.
\newblock In: {\sl \bibinfo{booktitle}{FSTTCS}}, \bibinfo{address}{Bangalore,
  India}.

\bibitemdeclare{inproceedings}{Dimitrova_Finkbeiner_12}
\bibitem{Dimitrova_Finkbeiner_12}
\bibinfo{author}{Rayna \surnamestart Dimitrova\surnameend} \&
  \bibinfo{author}{Bernd \surnamestart Finkbeiner\surnameend}
  (\bibinfo{year}{2012}): \emph{\bibinfo{title}{Counterexample-Guided Synthesis
  of Observation Predicates}}.
\newblock In: {\sl \bibinfo{booktitle}{FORMATS}}, \bibinfo{address}{London,
  UK}, pp. \bibinfo{pages}{107--122}.
\doi{10.1007/978-3-642-33365-1\_9}.

\bibitemdeclare{inproceedings}{Ehlers_10}
\bibitem{Ehlers_10}
\bibinfo{author}{R\"{u}diger \surnamestart Ehlers\surnameend}
  (\bibinfo{year}{2010}): \emph{\bibinfo{title}{Symbolic Bounded Synthesis}}.
\newblock In: {\sl \bibinfo{booktitle}{CAV}}, \bibinfo{address}{Edinburgh, UK}.
\doi{10.1007/978-3-642-14295-6\_33}.

\bibitemdeclare{inproceedings}{Ehlers_KH_12}
\bibitem{Ehlers_KH_12}
\bibinfo{author}{R{\"{u}}diger \surnamestart Ehlers\surnameend},
  \bibinfo{author}{Robert \surnamestart K{\"{o}}nighofer\surnameend} \&
  \bibinfo{author}{Georg \surnamestart Hofferek\surnameend}
  (\bibinfo{year}{2012}): \emph{\bibinfo{title}{Symbolically synthesizing small
  circuits}}.
\newblock In: {\sl \bibinfo{booktitle}{FMCAD}}, \bibinfo{address}{Cambridge,
  UK}, pp. \bibinfo{pages}{91--100}.


\bibitemdeclare{inproceedings}{Flanagan_Qadeer_02}
\bibitem{Flanagan_Qadeer_02}
\bibinfo{author}{Cormac \surnamestart Flanagan\surnameend} \&
  \bibinfo{author}{Shaz \surnamestart Qadeer\surnameend}
  (\bibinfo{year}{2002}): \emph{\bibinfo{title}{Predicate Abstraction for
  Software Verification}}.
\newblock In: {\sl \bibinfo{booktitle}{29th ACM SIGPLAN-SIGACT Symposium on
  Principles of Programming Languages}}, \bibinfo{address}{Portland, Oregon},
  pp. \bibinfo{pages}{191--202}.
\doi{10.1145/503272.503291}.

\bibitemdeclare{inproceedings}{Ganapathi_GP_06}
\bibitem{Ganapathi_GP_06}
\bibinfo{author}{Archana \surnamestart Ganapathi\surnameend},
  \bibinfo{author}{Viji \surnamestart Ganapathi\surnameend} \&
  \bibinfo{author}{David \surnamestart Patterson\surnameend}
  (\bibinfo{year}{2006}): \emph{\bibinfo{title}{{Windows XP} Kernel Crash
  Analysis}}.
\newblock In: {\sl \bibinfo{booktitle}{20th USENIX Large Installation System
  Administration Conference}}, \bibinfo{address}{Washington, DC, USA}, pp.
  \bibinfo{pages}{101--111}.


\bibitemdeclare{book}{Grumberg_Veith_08}
\bibitem{Grumberg_Veith_08}
\bibinfo{editor}{Orna \surnamestart Grumberg\surnameend} \&
  \bibinfo{editor}{Helmut \surnamestart Veith\surnameend}, editors
  (\bibinfo{year}{2008}): \emph{\bibinfo{title}{25 Years of Model Checking:
  History, Achievements, Perspectives}}.
\newblock \bibinfo{publisher}{Springer-Verlag}, \bibinfo{address}{Berlin,
  Heidelberg}.
\doi{10.1007/978-3-540-69850-0}.

\bibitemdeclare{inproceedings}{Henzinger_JM_03}
\bibitem{Henzinger_JM_03}
\bibinfo{author}{Thomas~A. \surnamestart Henzinger\surnameend},
  \bibinfo{author}{Ranjit \surnamestart Jhala\surnameend} \&
  \bibinfo{author}{Rupak \surnamestart Majumdar\surnameend}
  (\bibinfo{year}{2003}): \emph{\bibinfo{title}{Counterexample-guided
  control}}.
\newblock In: {\sl \bibinfo{booktitle}{ICALP}}, \bibinfo{address}{Eindhoven,
  The Netherlands}, pp. \bibinfo{pages}{886--902}.
\doi{10.1007/3-540-45061-0\_69}.

\bibitemdeclare{misc}{cofluent}
\bibitem{cofluent}
\bibinfo{author}{\surnamestart {Intel Corporation}\surnameend}:
  \emph{\bibinfo{title}{CoFluent Technolofy}}.
\newblock
  \bibinfo{howpublished}{\url{http://www.intel.com/content/www/us/en/cofluent/cofluent-difference.html}}.

\bibitemdeclare{article}{Jacobs_14}
\bibitem{Jacobs_14}
\bibinfo{author}{Swen \surnamestart Jacobs\surnameend} (\bibinfo{year}{2014}):
  \emph{\bibinfo{title}{Extended {AIGER} Format for Synthesis}}.
\newblock {\sl \bibinfo{journal}{CoRR}} \bibinfo{volume}{abs/1405.5793}.


\bibitemdeclare{inproceedings}{Jobstmann_Bloem_06}
\bibitem{Jobstmann_Bloem_06}
\bibinfo{author}{Barbara \surnamestart Jobstmann\surnameend} \&
  \bibinfo{author}{Roderick \surnamestart Bloem\surnameend}
  (\bibinfo{year}{2006}): \emph{\bibinfo{title}{Optimizations for {LTL}
  Synthesis}}.
\newblock In: {\sl \bibinfo{booktitle}{FMCAD}}, \bibinfo{address}{San Jose, CA,
  USA}, pp. \bibinfo{pages}{117--124}.
\doi{10.1109/fmcad.2006.22}.

\bibitemdeclare{inproceedings}{Jobstmann_GWB_07}
\bibitem{Jobstmann_GWB_07}
\bibinfo{author}{Barbara \surnamestart Jobstmann\surnameend},
  \bibinfo{author}{Stefan~J. \surnamestart Galler\surnameend},
  \bibinfo{author}{Martin \surnamestart Weiglhofer\surnameend} \&
  \bibinfo{author}{Roderick \surnamestart Bloem\surnameend}
  (\bibinfo{year}{2007}): \emph{\bibinfo{title}{Anzu: {A} Tool for Property
  Synthesis}}.
\newblock In: {\sl \bibinfo{booktitle}{CAV}}, \bibinfo{address}{Berlin,
  Germany}, pp. \bibinfo{pages}{258--262}.
\doi{10.1007/978-3-540-73368-3\_29}.

\bibitemdeclare{inproceedings}{Kadav_RS_09}
\bibitem{Kadav_RS_09}
\bibinfo{author}{Asim \surnamestart Kadav\surnameend},
  \bibinfo{author}{Matthew~J. \surnamestart Renzelmann\surnameend} \&
  \bibinfo{author}{Michael~M. \surnamestart Swift\surnameend}
  (\bibinfo{year}{2009}): \emph{\bibinfo{title}{Tolerating Hardware Device
  Failures in Software}}.
\newblock In: {\sl \bibinfo{booktitle}{22nd ACM Symposium on Operating Systems
  Principles}}, \bibinfo{address}{Big Sky, MT, USA}.
\doi{10.1145/1629575.1629582}.

\bibitemdeclare{inproceedings}{Klein_EHACDEEKNSTW_09}
\bibitem{Klein_EHACDEEKNSTW_09}
\bibinfo{author}{Gerwin \surnamestart Klein\surnameend}, \bibinfo{author}{Kevin
  \surnamestart Elphinstone\surnameend}, \bibinfo{author}{Gernot \surnamestart
  Heiser\surnameend}, \bibinfo{author}{June \surnamestart
  Andronick\surnameend}, \bibinfo{author}{David \surnamestart Cock\surnameend},
  \bibinfo{author}{Philip \surnamestart Derrin\surnameend},
  \bibinfo{author}{Dhammika \surnamestart Elkaduwe\surnameend},
  \bibinfo{author}{Kai \surnamestart Engelhardt\surnameend},
  \bibinfo{author}{Rafal \surnamestart Kolanski\surnameend},
  \bibinfo{author}{Michael \surnamestart Norrish\surnameend},
  \bibinfo{author}{Thomas \surnamestart Sewell\surnameend},
  \bibinfo{author}{Harvey \surnamestart Tuch\surnameend} \&
  \bibinfo{author}{Simon \surnamestart Winwood\surnameend}
  (\bibinfo{year}{2009}): \emph{\bibinfo{title}{{seL4}: Formal Verification of
  an {OS} Kernel}}.
\newblock In: {\sl \bibinfo{booktitle}{22nd {ACM} Symposium on Operating
  Systems Principles}}, \bibinfo{address}{Big Sky, MT, USA}, pp.
  \bibinfo{pages}{207--220}.
\doi{10.1145/1629575.1629596}.

\bibitemdeclare{article}{Kneuss_KKS_13}
\bibitem{Kneuss_KKS_13}
\bibinfo{author}{Etienne \surnamestart Kneuss\surnameend},
  \bibinfo{author}{Viktor \surnamestart Kuncak\surnameend},
  \bibinfo{author}{Ivan \surnamestart Kuraj\surnameend} \&
  \bibinfo{author}{Philippe \surnamestart Suter\surnameend}
  (\bibinfo{year}{2013}): \emph{\bibinfo{title}{On Integrating Deductive
  Synthesis and Verification Systems}}.
\newblock {\sl \bibinfo{journal}{CoRR}} \bibinfo{volume}{abs/1304.5661}.


\bibitemdeclare{inproceedings}{Koninghofer_HB_09}
\bibitem{Koninghofer_HB_09}
\bibinfo{author}{Robert \surnamestart K{\"{o}}nighofer\surnameend},
  \bibinfo{author}{Georg \surnamestart Hofferek\surnameend} \&
  \bibinfo{author}{Roderick \surnamestart Bloem\surnameend}
  (\bibinfo{year}{2009}): \emph{\bibinfo{title}{Debugging formal specifications
  using simple counterstrategies}}.
\newblock In: {\sl \bibinfo{booktitle}{FMCAD}}, \bibinfo{address}{Austin,
  Texas, USA}, pp. \bibinfo{pages}{152--159}.
\doi{10.1109/fmcad.2009.5351127}.

\bibitemdeclare{article}{Konighofer_HB_13}
\bibitem{Konighofer_HB_13}
\bibinfo{author}{Robert \surnamestart K{\"{o}}nighofer\surnameend},
  \bibinfo{author}{Georg \surnamestart Hofferek\surnameend} \&
  \bibinfo{author}{Roderick \surnamestart Bloem\surnameend}
  (\bibinfo{year}{2013}): \emph{\bibinfo{title}{Debugging formal
  specifications: a practical approach using model-based diagnosis and
  counterstrategies}}.
\newblock {\sl \bibinfo{journal}{STTT}}
  \bibinfo{volume}{15}(\bibinfo{number}{5-6}), pp. \bibinfo{pages}{563--583}.
\doi{10.1007/s10009-011-0221-y}.

\bibitemdeclare{incollection}{Kupferman_Vardi_2000}
\bibitem{Kupferman_Vardi_2000}
\bibinfo{author}{Orna \surnamestart Kupferman\surnameend} \&
  \bibinfo{author}{Moshe \surnamestart Vardi\surnameend}
  (\bibinfo{year}{2000}): \emph{\bibinfo{title}{Synthesis with Incomplete
  Information}}.
\newblock In: {\sl \bibinfo{booktitle}{Advances in Temporal Logic}},
  \bibinfo{volume}{16}, \bibinfo{publisher}{Springer Netherlands}, pp.
  \bibinfo{pages}{109--127}.
\doi{10.1007/978-94-015-9586-5\_6}.

\bibitemdeclare{article}{Leslie_CFGGMPSEH_05}
\bibitem{Leslie_CFGGMPSEH_05}
\bibinfo{author}{Ben \surnamestart Leslie\surnameend}, \bibinfo{author}{Peter
  \surnamestart Chubb\surnameend}, \bibinfo{author}{Nicholas \surnamestart
  FitzRoy-Dale\surnameend}, \bibinfo{author}{Stefan \surnamestart
  G\"{o}tz\surnameend}, \bibinfo{author}{Charles \surnamestart
  Gray\surnameend}, \bibinfo{author}{Luke \surnamestart Macpherson\surnameend},
  \bibinfo{author}{Daniel \surnamestart Potts\surnameend},
  \bibinfo{author}{Yueting~(Rita) \surnamestart Shen\surnameend},
  \bibinfo{author}{Kevin \surnamestart Elphinstone\surnameend} \&
  \bibinfo{author}{Gernot \surnamestart Heiser\surnameend}
  (\bibinfo{year}{2005}): \emph{\bibinfo{title}{User-level Device Drivers:
  Achieved Performance}}.
\newblock {\sl \bibinfo{journal}{Journal of Computer Science and Technology}}
  \bibinfo{volume}{20}(\bibinfo{number}{5}), pp. \bibinfo{pages}{654--664}.
\doi{10.1007/s11390-005-0654-4}.

\bibitemdeclare{inproceedings}{LeVasseur_USG_04}
\bibitem{LeVasseur_USG_04}
\bibinfo{author}{Joshua \surnamestart LeVasseur\surnameend},
  \bibinfo{author}{Volkmar \surnamestart Uhlig\surnameend},
  \bibinfo{author}{Jan \surnamestart Stoess\surnameend} \&
  \bibinfo{author}{Stefan \surnamestart G\"{o}tz\surnameend}
  (\bibinfo{year}{2004}): \emph{\bibinfo{title}{Unmodified Device Driver Reuse
  and Improved System Dependability via Virtual Machines}}.
\newblock In: {\sl \bibinfo{booktitle}{6th Symposium on Operating Systems
  Design and Implementation}}, \bibinfo{address}{San Francisco, CA, USA}, pp.
  \bibinfo{pages}{17--30}.


\bibitemdeclare{inproceedings}{Merillon_RCMM_00}
\bibitem{Merillon_RCMM_00}
\bibinfo{author}{Fabrice \surnamestart M\'{e}rillon\surnameend},
  \bibinfo{author}{Laurent \surnamestart R\'{e}veill\`{e}re\surnameend},
  \bibinfo{author}{Charles \surnamestart Consel\surnameend},
  \bibinfo{author}{Renaud \surnamestart Marlet\surnameend} \&
  \bibinfo{author}{Gilles \surnamestart Muller\surnameend}
  (\bibinfo{year}{2000}): \emph{\bibinfo{title}{Devil: An {IDL} for hardware
  programming}}.
\newblock In: {\sl \bibinfo{booktitle}{4th USENIX Symposium on Operating
  Systems Design and Implementation}}, \bibinfo{address}{San Diego, CA, USA},
  pp. \bibinfo{pages}{17--30}.


\bibitemdeclare{inproceedings}{Morgenstern_GS_13}
\bibitem{Morgenstern_GS_13}
\bibinfo{author}{Andreas \surnamestart Morgenstern\surnameend},
  \bibinfo{author}{Manuel \surnamestart Gesell\surnameend} \&
  \bibinfo{author}{Klaus \surnamestart Schneider\surnameend}
  (\bibinfo{year}{2013}): \emph{\bibinfo{title}{Solving Games Using Incremental
  Induction}}.
\newblock In: {\sl \bibinfo{booktitle}{IFM}}, \bibinfo{address}{Turku,
  Finland}, pp. \bibinfo{pages}{177--191}.
\doi{10.1007/978-3-642-38613-8\_13}.

\bibitemdeclare{inproceedings}{Moskewicz_MZZM_01}
\bibitem{Moskewicz_MZZM_01}
\bibinfo{author}{Matthew~W. \surnamestart Moskewicz\surnameend},
  \bibinfo{author}{Conor~F. \surnamestart Madigan\surnameend},
  \bibinfo{author}{Ying \surnamestart Zhao\surnameend}, \bibinfo{author}{Lintao
  \surnamestart Zhang\surnameend} \& \bibinfo{author}{Sharad \surnamestart
  Malik\surnameend} (\bibinfo{year}{2001}): \emph{\bibinfo{title}{Chaff:
  engineering an efficient {SAT} solver}}.
\newblock In: {\sl \bibinfo{booktitle}{DAC}}, \bibinfo{address}{Las Vegas, NV,
  USA}, pp. \bibinfo{pages}{530--535}.
\doi{10.1109/dac.2001.935565}.

\bibitemdeclare{inproceedings}{Narodytska_LBRW_14}
\bibitem{Narodytska_LBRW_14}
\bibinfo{author}{Nina \surnamestart Narodytska\surnameend},
  \bibinfo{author}{Alexander \surnamestart Legg\surnameend},
  \bibinfo{author}{Fahiem \surnamestart Bacchus\surnameend},
  \bibinfo{author}{Leonid \surnamestart Ryzhyk\surnameend} \&
  \bibinfo{author}{Adam \surnamestart Walker\surnameend}
  (\bibinfo{year}{2014}): \emph{\bibinfo{title}{Solving Games without
  Controllable Predecessor}}.
\newblock In: {\sl \bibinfo{booktitle}{CAV}}, \bibinfo{address}{Vienna,
  Austria}.
\doi{10.1007/978-3-319-08867-9\_35}.

\bibitemdeclare{inproceedings}{Naujokat_LS_12}
\bibitem{Naujokat_LS_12}
\bibinfo{author}{Stefan \surnamestart Naujokat\surnameend},
  \bibinfo{author}{Anna-Lena \surnamestart Lamprecht\surnameend} \&
  \bibinfo{author}{Bernhard \surnamestart Steffen\surnameend}
  (\bibinfo{year}{2012}): \emph{\bibinfo{title}{Loose Programming with
  {PROPHETS}}}.
\newblock In: {\sl \bibinfo{booktitle}{FASE'12}}, \bibinfo{address}{Tallinn,
  Estonia}, pp. \bibinfo{pages}{94--98}.
\doi{10.1007/978-3-642-28872-2\_7}.

\bibitemdeclare{inproceedings}{ONils_OJ_98}
\bibitem{ONils_OJ_98}
\bibinfo{author}{Mattias \surnamestart O'Nils\surnameend},
  \bibinfo{author}{Johnny \surnamestart \"{O}berg\surnameend} \&
  \bibinfo{author}{Axel \surnamestart Jantsch\surnameend}
  (\bibinfo{year}{1998}): \emph{\bibinfo{title}{Grammar Based Modelling and
  Synthesis of Device Drivers and Bus Interfaces}}.
\newblock \bibinfo{address}{Washington, DC, USA}.
\doi{10.1109/EURMIC.1998.711776}.

\bibitemdeclare{misc}{uart}
\bibitem{uart}
\emph{\bibinfo{title}{{16550 UART core}}}.
\newblock
  \bibinfo{howpublished}{\url{http://opencores.org/project,a_vhd_16550_uart}}.

\bibitemdeclare{inproceedings}{Palix_GSCLM_11}
\bibitem{Palix_GSCLM_11}
\bibinfo{author}{Nicolas \surnamestart Palix\surnameend},
  \bibinfo{author}{Ga\"{e}l \surnamestart Thomas\surnameend},
  \bibinfo{author}{Suman \surnamestart Saha\surnameend},
  \bibinfo{author}{Christophe \surnamestart Calv\`{e}s\surnameend},
  \bibinfo{author}{Julia \surnamestart Lawall\surnameend} \&
  \bibinfo{author}{Gilles \surnamestart Muller\surnameend}
  (\bibinfo{year}{2011}): \emph{\bibinfo{title}{Faults in {Linux}: ten years
  later}}.
\newblock In: {\sl \bibinfo{booktitle}{16th International Conference on
  Architectural Support for Programming Languages and Operating Systems}},
  \bibinfo{address}{Newport Beach, CA, USA}, pp. \bibinfo{pages}{305--318}.
\doi{10.1145/1950365.1950401}.

\bibitemdeclare{inproceedings}{Piterman_PS_06}
\bibitem{Piterman_PS_06}
\bibinfo{author}{Nir \surnamestart Piterman\surnameend}, \bibinfo{author}{Amir
  \surnamestart Pnueli\surnameend} \& \bibinfo{author}{Yaniv \surnamestart
  Sa'ar\surnameend} (\bibinfo{year}{2006}): \emph{\bibinfo{title}{Synthesis of
  {Reactive}(1) designs}}.
\newblock In: {\sl \bibinfo{booktitle}{7th International Conference on
  Verification, Model Checking and Abstract Interpretation}}, pp.
  \bibinfo{pages}{364--380}.
\doi{10.1007/11609773\_24}.

\bibitemdeclare{inproceedings}{Pnueli_Rosner_89}
\bibitem{Pnueli_Rosner_89}
\bibinfo{author}{Amir \surnamestart Pnueli\surnameend} \& \bibinfo{author}{Roni
  \surnamestart Rosner\surnameend} (\bibinfo{year}{1989}):
  \emph{\bibinfo{title}{On the Synthesis of a Reactive Module}}.
\newblock In: {\sl \bibinfo{booktitle}{POPL}}, \bibinfo{address}{Austin, Texas,
  USA}, pp. \bibinfo{pages}{179--190}.
\doi{10.1145/75277.75293}.

\bibitemdeclare{article}{Raskin_CDH_07}
\bibitem{Raskin_CDH_07}
\bibinfo{author}{Jean{-}Fran{\c{c}}ois \surnamestart Raskin\surnameend},
  \bibinfo{author}{Krishnendu \surnamestart Chatterjee\surnameend},
  \bibinfo{author}{Laurent \surnamestart Doyen\surnameend} \&
  \bibinfo{author}{Thomas~A. \surnamestart Henzinger\surnameend}
  (\bibinfo{year}{2007}): \emph{\bibinfo{title}{Algorithms for Omega-Regular
  Games with Imperfect Information}}.
\newblock {\sl \bibinfo{journal}{Logical Methods in Computer Science}}
  \bibinfo{volume}{3}(\bibinfo{number}{3}).
\doi{10.2168/LMCS-3(3:4)2007}.

\bibitemdeclare{inproceedings}{Renzelmann_Swift_09}
\bibitem{Renzelmann_Swift_09}
\bibinfo{author}{Matthew~J. \surnamestart Renzelmann\surnameend} \&
  \bibinfo{author}{Michael~M. \surnamestart Swift\surnameend}
  (\bibinfo{year}{2009}): \emph{\bibinfo{title}{{Decaf}: Moving Device Drivers
  to a Moderm Language}}.
\newblock In: {\sl \bibinfo{booktitle}{USENIX Annual Technical Conference}},
  \bibinfo{address}{San Diego, CA, USA}.


\bibitemdeclare{inproceedings}{Rudell_1993}
\bibitem{Rudell_1993}
\bibinfo{author}{Richard \surnamestart Rudell\surnameend}
  (\bibinfo{year}{1993}): \emph{\bibinfo{title}{Dynamic variable ordering for
  ordered binary decision diagrams}}.
\newblock In: {\sl \bibinfo{booktitle}{ICCAD}}, \bibinfo{address}{Santa Clara,
  CA, USA}, pp. \bibinfo{pages}{42--47}.
\doi{10.1007/978-1-4615-0292-0\_5}.

\bibitemdeclare{misc}{tsl}
\bibitem{tsl}
\bibinfo{author}{Leonid \surnamestart Ryzhyk\surnameend}
  (\bibinfo{year}{2014}): \emph{\bibinfo{title}{{TSL2} Reference Manual}}.

\bibitemdeclare{inproceedings}{Ryzhyk_CKSH_09}
\bibitem{Ryzhyk_CKSH_09}
\bibinfo{author}{Leonid \surnamestart Ryzhyk\surnameend},
  \bibinfo{author}{Peter \surnamestart Chubb\surnameend}, \bibinfo{author}{Ihor
  \surnamestart Kuz\surnameend}, \bibinfo{author}{Etienne \surnamestart
  Le~Sueur\surnameend} \& \bibinfo{author}{Gernot \surnamestart
  Heiser\surnameend} (\bibinfo{year}{2009}): \emph{\bibinfo{title}{Automatic
  Device Driver Synthesis with {Termite}}}.
\newblock In: {\sl \bibinfo{booktitle}{22nd ACM Symposium on Operating Systems
  Principles}}, \bibinfo{address}{Big Sky, MT, USA}.
\doi{10.1145/1629575.1629583}.

\bibitemdeclare{inproceedings}{Ryzhyk_KMRVH_11}
\bibitem{Ryzhyk_KMRVH_11}
\bibinfo{author}{Leonid \surnamestart Ryzhyk\surnameend}, \bibinfo{author}{John
  \surnamestart Keys\surnameend}, \bibinfo{author}{Balachandra \surnamestart
  Mirla\surnameend}, \bibinfo{author}{Arun \surnamestart Raghunath\surnameend},
  \bibinfo{author}{Mona \surnamestart Vij\surnameend} \&
  \bibinfo{author}{Gernot \surnamestart Heiser\surnameend}
  (\bibinfo{year}{2011}): \emph{\bibinfo{title}{Improved Device Driver
  Reliability Through Hardware Verification Reuse}}.
\newblock In: {\sl \bibinfo{booktitle}{16th International Conference on
  Architectural Support for Programming Languages and Operating Systems}},
  \bibinfo{address}{Newport Beach, CA, USA}.
\doi{10.1145/1950365.1950383}.

\bibitemdeclare{inproceedings}{Ryzhyk_WKLRSV_14}
\bibitem{Ryzhyk_WKLRSV_14}
\bibinfo{author}{Leonid \surnamestart Ryzhyk\surnameend}, \bibinfo{author}{Adam
  \surnamestart Walker\surnameend}, \bibinfo{author}{John \surnamestart
  Keys\surnameend}, \bibinfo{author}{Alexander \surnamestart Legg\surnameend},
  \bibinfo{author}{Arun \surnamestart Raghunath\surnameend},
  \bibinfo{author}{Michael \surnamestart Stumm\surnameend} \&
  \bibinfo{author}{Mona \surnamestart Vij\surnameend} (\bibinfo{year}{2014}):
  \emph{\bibinfo{title}{User-Guided Device Driver Synthesis}}.
\newblock In: {\sl \bibinfo{booktitle}{OSDI}}, \bibinfo{address}{Broomfield,
  CO, USA}.

\bibitemdeclare{book}{Sipser_96}
\bibitem{Sipser_96}
\bibinfo{author}{Michael \surnamestart Sipser\surnameend}
  (\bibinfo{year}{1996}): \emph{\bibinfo{title}{Introduction to the Theory of
  Computation}}, \bibinfo{edition}{1st} edition.
\newblock \bibinfo{publisher}{International Thomson Publishing}.
\doi{10.1145/230514.571645}.

\bibitemdeclare{inproceedings}{Solar-Lezama_RBE_05}
\bibitem{Solar-Lezama_RBE_05}
\bibinfo{author}{Armando \surnamestart Solar-Lezama\surnameend},
  \bibinfo{author}{Rodric \surnamestart Rabbah\surnameend},
  \bibinfo{author}{Rastislav \surnamestart Bod\'{\i}k\surnameend} \&
  \bibinfo{author}{Kemal \surnamestart Ebcio\u{g}lu\surnameend}
  (\bibinfo{year}{2005}): \emph{\bibinfo{title}{Programming by Sketching for
  Bit-streaming Programs}}.
\newblock In: {\sl \bibinfo{booktitle}{PLDI}}, \bibinfo{address}{Chicago,
  Illinois, USA}.
\doi{10.1145/1064978.1065045}.

\bibitemdeclare{inproceedings}{Spear_RHHL_06}
\bibitem{Spear_RHHL_06}
\bibinfo{author}{Michael~F. \surnamestart Spear\surnameend},
  \bibinfo{author}{Tom \surnamestart Roeder\surnameend}, \bibinfo{author}{Orion
  \surnamestart Hodson\surnameend}, \bibinfo{author}{Galen~C. \surnamestart
  Hunt\surnameend} \& \bibinfo{author}{Steven \surnamestart Levi\surnameend}
  (\bibinfo{year}{2006}): \emph{\bibinfo{title}{Solving the starting problem:
  device drivers as self-describing artifacts}}.
\newblock In: {\sl \bibinfo{booktitle}{1st EuroSys Conference}},
  \bibinfo{address}{Leuven, Belgium}, pp. \bibinfo{pages}{45--57}.
\doi{10.1145/1217935.1217941}.

\bibitemdeclare{inproceedings}{Swift_BL_03}
\bibitem{Swift_BL_03}
\bibinfo{author}{Michael~M. \surnamestart Swift\surnameend},
  \bibinfo{author}{Brian~N. \surnamestart Bershad\surnameend} \&
  \bibinfo{author}{Henry~M. \surnamestart Levy\surnameend}
  (\bibinfo{year}{2003}): \emph{\bibinfo{title}{Improving the Reliability of
  Commodity Operating Systems}}.
\newblock In: {\sl \bibinfo{booktitle}{19th ACM Symposium on Operating Systems
  Principles}}, \bibinfo{address}{Bolton Landing (Lake George), New York, USA}.
\doi{10.1145/1165389.945466}.

\bibitemdeclare{misc}{vp}
\bibitem{vp}
\bibinfo{author}{\surnamestart {Synopsys}\surnameend}:
  \emph{\bibinfo{title}{Virtual Prototyping Models}}.
\newblock
  \bibinfo{howpublished}{\url{http://www.synopsys.com/Systems/VirtualPrototyping/VPModels}}.

\bibitemdeclare{inproceedings}{Thomas_95}
\bibitem{Thomas_95}
\bibinfo{author}{Wolfgang \surnamestart Thomas\surnameend}
  (\bibinfo{year}{1995}): \emph{\bibinfo{title}{On the Synthesis of Strategies
  in Infinite Games}}.
\newblock In: {\sl \bibinfo{booktitle}{12th Annual Symposium on Theoretical
  Aspects of Computer Science}}, pp. \bibinfo{pages}{1--13}.
\doi{10.1007/3-540-59042-0\_57}.

\bibitemdeclare{misc}{ddgen}
\bibitem{ddgen}
\bibinfo{author}{\surnamestart {Vayavya Labs}\surnameend}:
  \emph{\bibinfo{title}{{Device Driver Generator tool}}}.
\newblock \bibinfo{howpublished}{\url{http://vayavyalabs.com/products/}}.

\bibitemdeclare{inproceedings}{Walker_Ryzhyk_14}
\bibitem{Walker_Ryzhyk_14}
\bibinfo{author}{Adam \surnamestart Walker\surnameend} \&
  \bibinfo{author}{Leonid \surnamestart Ryzhyk\surnameend}
  (\bibinfo{year}{2014}): \emph{\bibinfo{title}{Predicate Abstraction for
  Reactive Synthesis}}.
\newblock In: {\sl \bibinfo{booktitle}{FMCAD}}, \bibinfo{address}{Lausanne,
  Switzerland}.
\doi{10.1109/FMCAD.2014.6987617}.

\bibitemdeclare{misc}{dml}
\bibitem{dml}
\bibinfo{author}{\surnamestart {Wind River}\surnameend} (\bibinfo{year}{2010}):
  \emph{\bibinfo{title}{{Wind River Simics Model Builder} reference manual.
  Version 4.4}}.

\bibitemdeclare{misc}{dml_ug}
\bibitem{dml_ug}
\bibinfo{author}{\surnamestart {Wind River}\surnameend} (\bibinfo{year}{2010}):
  \emph{\bibinfo{title}{{Wind River Simics Model Builder} user guide. Version
  4.4}}.

\bibitemdeclare{misc}{ds12887}
\bibitem{ds12887}
\emph{\bibinfo{title}{{WindRiver Simics DS12887 Model}}}.
\newblock
  \bibinfo{howpublished}{\url{http://www.windriver.com/products/simics}}.

\bibitemdeclare{misc}{Yavatkar_12}
\bibitem{Yavatkar_12}
\bibinfo{author}{Raj \surnamestart Yavatkar\surnameend} (\bibinfo{year}{2012}):
  \emph{\bibinfo{title}{Era of {SoCs}, presentation at the {Intel Workshop on
  Device Driver Reliability, Modeling and Synthesis}}}.

\bibitemdeclare{inproceedings}{Zhou_CABE_06}
\bibitem{Zhou_CABE_06}
\bibinfo{author}{Feng \surnamestart Zhou\surnameend}, \bibinfo{author}{Jeremy
  \surnamestart Condit\surnameend}, \bibinfo{author}{Zachary \surnamestart
  Anderson\surnameend}, \bibinfo{author}{Ilya \surnamestart Bagrak\surnameend},
  \bibinfo{author}{Rob \surnamestart Ennals\surnameend},
  \bibinfo{author}{Matthew \surnamestart Harren\surnameend},
  \bibinfo{author}{George \surnamestart Necula\surnameend} \&
  \bibinfo{author}{Eric \surnamestart Brewer\surnameend}
  (\bibinfo{year}{2006}): \emph{\bibinfo{title}{{SafeDrive}: Safe and
  Recoverable Extensions Using Language-Based Techniques}}.
\newblock In: {\sl \bibinfo{booktitle}{7th USENIX Symposium on Operating
  Systems Design and Implementation}}, \bibinfo{address}{Seattle, WA, USA}, pp.
  \bibinfo{pages}{45--60}.


\end{thebibliography}

\end{document}